\documentclass[a4paper,12pt,german,american]{book} 

\usepackage{makeidx}
\usepackage{acronym}
\usepackage{epsfig}
\usepackage{subfigure}

\usepackage{amsmath}
\usepackage{amssymb}
\usepackage{cite} 
\usepackage{array}
\usepackage{wrapfig}

\usepackage[breaklinks]{hyperref} 

\usepackage[german,american]{babel}
\usepackage[latin1]{inputenc}

\usepackage{fancyhdr}
\fancypagestyle{plain}{%
  \fancyhf{}                       
  \cfoot{\thepage}                 
  
  }

\newcommand{\Ix}[1]{#1\index{#1}}

\newcommand{\beqs}{\begin{equation*}}
\newcommand{\beq}{\begin{equation}}

\newcommand{\eeqs}{\end{equation*}}
\newcommand{\eeq}{\end{equation}}

\newcommand{\beqas}{\begin{eqnarray*}}
\newcommand{\beqa}{\begin{eqnarray}}

\newcommand{\eeqas}{\end{eqnarray*}}
\newcommand{\eeqa}{\end{eqnarray}}

\newcommand{\eq}[2]{\begin{equation} #1 \label{#2} \end{equation}}

\newcommand{\eps}{\varepsilon}
\newcommand{\al}{\alpha}
\newcommand{\be}{\beta}
\newcommand{\ga}{\gamma}
\newcommand{\de}{\delta}
\newcommand{\om}{\omega}
\newcommand{\ka}{\kappa}
\newcommand{\la}{\lambda}

\newcommand{\Ga}{\Gamma}

\newcommand{\Om}{\Omega}

\newcommand{\La}{\Lambda}
\newcommand{\Si}{\Sigma}

\newcommand{\blist}{\begin{itemize}}

\newcommand{\elist}{\end{itemize}}

\providecommand{\href}[2]{#2}

\newcommand{\chapquote}[2]{{\it #1}\begin{flushright}#2\end{flushright}}

\newcommand{\clearemptydoublepage}{\newpage{\pagestyle{empty}\cleardoublepage}}
\newcommand{\clearplaindoublepage}{\newpage{\pagestyle{plain}\cleardoublepage}}

\DeclareFontFamily{OT1}{rsfs}{}
\DeclareFontShape{OT1}{rsfs}{m}{n}{ <-7> rsfs5 <7-10> rsfs7 <10->rsfs10}{} 
\DeclareMathAlphabet{\mycal}{OT1}{rsfs}{m}{n}

\newcommand{\plr}[1]{\overleftrightarrow{\partial_{#1}}}

\newcommand{\lrpd}[1]{\overleftrightarrow{\partial_{#1}}}

\newcommand{\psib}{\overline{\psi}}
\newcommand{\chib}{\overline{\chi}}
\newcommand{\chid}{\chi^*}
\newcommand{\chitb}{\overline{\tilde{\chi}}}
\newcommand{\chit}{\tilde{\chi}}
\newcommand{\etatb}{\overline{\tilde{\eta}}}
\newcommand{\etat}{\tilde{\eta}}
\newcommand{\complexc}{{\mathbb C}}
\def\etab{\overline{\eta}}

\usepackage{slashed}
 
\def\Det{{\rm Det}}
\def\Dir{\slashed{D}}

\def\Zt{\tilde{Z}}
\def\Qh{\hat{Q}}
\def\Bh{\hat{B}}

\def\ol{\overline}
\def\cV{{\cal V}}
\def\cL{{\cal L}}
\def\cH{{\cal H}}
\def\cD{{\cal D}}
\def\cN{{\cal N}}
\def\cM{{\cal M}}

\def\cR{{\cal R}}
\def\cC{{\cal C}}
\def\cO{{\cal O}}
\def\cA{{\cal A}}

\def\extd{{\rm d}}

\newcommand{\pd}[2]{\frac{\partial {#1}}{\partial {#2}}}

\newcommand{\pld}[2]{\frac{\partial^L #1}{\partial #2}}

\newcommand{\td}[2]{\frac{\mathrm{d} #1}{\mathrm{d} #2}}

\newcommand{\poiss}[2]{\{ #1 , #2 \}}
\newcommand{\spoiss}[2]{ \{ #1 , #2 \}}
\newcommand{\dirac}[2]{\{ #1 , #2 \}^*}

\def\sdet{{\rm sdet}}
\def\str{{\rm str}}
\def\tr{{\rm tr}}
\def\Tr{{\rm Tr}_{L^2}}

\def\atbdry{\Big|_{\partial \cM}}
\def\atbdry0{\Big|_{\partial \cM_0}}
\def\atbdry1{\Big|_{\partial \cM_1}}

\newlength{\oldoddmargin}
\newlength{\oldevenmargin}
\def\act{S_{\rm 2DG}}
\def\gh{{\rm gh}}

\begin{document}

\ifx\href\undefined\else\hypersetup{linktocpage=true}\fi


\frontmatter

\fancyhf{}%

\selectlanguage{american}

\begin{titlepage}

\thispagestyle{empty}

    \centering

    \vspace*{2\baselineskip}

    {\huge
      \textbf{Diploma Thesis}}
        
    \vspace{3\baselineskip}

    {\Huge
      \textbf{Classical and Quantum Dilaton} \\
      \textbf{Gravity in Two Dimensions} \\
      \textbf{with Fermions}\\}
    
    \vspace{4\baselineskip}

    Presented to the Faculty for Physics and Geosciences \\
    University of Leipzig

        
    
    \vspace{1.5\baselineskip}
    
    By

    \vspace{1.5\baselineskip}
    
    \textbf{René Meyer} \\
     E-mail: \texttt{rene.meyer@itp.uni-leipzig.de}

    \vspace{5\baselineskip}
        
    Under the Supervision of \\
    Prof. Dr. Gerd Rudolph and Dr. Daniel Grumiller\\
    of the Institute for Theoretical Physics

    \vspace{\fill}
  
    \leftline{Leipzig, May $31^{\rm st}$, 2006.}

\newpage

\vspace*{10cm}

\thispagestyle{empty}

\end{titlepage}

\begin{samepage} 

\thispagestyle{plain}




\chapter*{Acknowledgments}

\pagenumbering{roman}

I am deeply indebted to Dr. Daniel Grumiller who, with much patience,
guided me in my scientific work while at the same time always having
an open ear for my questions. I also want to thank Prof. Dr. Gerd
Rudolph for taking responsibility for the official supervision of this
thesis, for friendly welcoming me in his research group and for the
interest he expressed towards the topic of this thesis. Furthermore, I
am grateful to Dr.  Dmitri Vassilevich for his constant scientific and
non-scientific advice and for many stimulating comments and
discussions.


My parents deserve every thank for the support they gave me in times
which were far from being easy for themselves, as well as my wife,
whose love is a constant source of strength and encouragement for me.
Thanks also go to all my friends for the good times we had during our
studies and for helping and supporting me in every manner.


Last but not least I am very grateful to the German National Academic
Foundation, whose scholarship enabled me to stay in Hangzhou (P.R.
China) during the academic year 2003/04 and made years of study
without monetary worries possible. In this regard I want to express my
gratitude once more towards Prof. Dr. Gerd Rudolph and Prof. Dr.
Klaus Sibold for supporting me in extending the scholarship.


\end{samepage}
\clearplaindoublepage

\begin{samepage}

\thispagestyle{plain}




\chapter*{Abstract}

In this thesis the first order formulation of generalized dilaton
gravities in two dimensions coupled to a Dirac fermion is considered.
After a Hamiltonian analysis of the gauge symmetries and constraints
of the theory and fixing Eddington-Finkelstein gauge by use of the
Batalin-Vilkovisky-Fradkin method, the system is quantized in the
Feynman path integral approach. It turns out that the path integral
over the dilaton gravity sector can be evaluated exactly, while in the
matter sector perturbative methods are applied. The gravitationally
induced four-fermi scattering vertices as well as asymptotic states
are calculated, and -- as for dilaton gravities coupled to scalar
fields -- a ``virtual black hole'' is found to form as an intermediary
geometric state in scattering processes. The results are compared to
the well-known scalar case, and evidence for bosonization in this
context is found.

\end{samepage}

\clearplaindoublepage

\begin{samepage}

\thispagestyle{plain}


\chapter*{}

\chapquote{Das ewig Unbegreifliche an der Welt ist ihre Begreiflichkeit.}{Werner Karl Heisenberg}

\end{samepage}

\clearplaindoublepage

\lhead[\thepage]{\slshape \contentsname}                       
\rhead[\slshape \contentsname]{\thepage}
\tableofcontents 

\clearemptydoublepage




\mainmatter


\lhead[\thepage]{\slshape \leftmark}
\rhead[\slshape \rightmark]{\thepage}
\renewcommand{\chaptermark}[1]{
  \markboth{\chaptername\ \thechapter.\ #1}{}}
\renewcommand{\sectionmark}[1]{
  \markright{#1}}

\chapter{Introduction}\label{ch:intro}

\chapquote{Gleichwohl müßten die Atome zufolge der inneratomischen
  Elektronenbe\-wegung nicht nur elektromagnetische, sondern auch
  Gravitationsenergie aus\-strahlen, wenn auch in winzigem Betrage. Da
  dies in Wahrheit in der Natur nicht zu\-treffen dürfte, so scheint es,
  daß die Quantentheorie nicht nur die Maxwellsche Elektrodynamik,
  sondern auch die neue Gravitationstheorie wird modifizieren müssen.}
{Albert Einstein \cite{Einstein:1916cc}}

Today, ninety years after Albert Einstein's discovery of General
Relativity \cite{Einstein:1915by,Einstein:1915ca} and eighty years
after the advent of quantum mechanics with the seminal papers by
Heisenberg and Schrödinger \cite{Heisenberg:1925,Schroedinger:1926},
an end of the search for a consistent and unique quantum theory of
gravity is still not in sight. Many different approaches (cf.  e.g.
the recent progress report \cite{Carlip:2001wq}) are available,
amongst which the most prominent ones are loop quantum gravity
and string theory%
, all stressing different aspects and succeeding in solving
different problems while at the same time suffering from different
drawbacks.  The failure of our efforts in quantizing gravity until now
can be traced back to conceptual and partly to technical issues.
Concerning the conceptual problems, the foundations of general
relativity, namely a dynamical space-time with no preferred reference
frame, seem to be incompatible with the needs of quantum theory, which
in its Hamiltonian formulation requires a choice of a preferred time
direction and thus violates general covariance. Also the notion of
causality in quantum field theory becomes meaningless if the metric of
the underlying space-time is subject to quantum fluctuations which so
to speak ``smear out'' the causal structure. On a deeper level,
quantum gravity requires the quantization of space-time itself, a
notion which first has to be given proper meaning by stating which the
valid observables are for such a quantum space-time. Technical
difficulties arise mostly from the already complicated and highly
nonlinear structure of general relativity. But if the classical
dynamics of gravity already is this complicated, one can not expect
quantization to be a much easier task.

One way to a better understanding of the conceptual issues behind
quantum gravity is to consider gravitation in lower dimensions to
simplify the dynamics of the theory. Experience with quantum field
theory on Minkowski space shows that models in two dimensions often
exhibit integrable or even topological structures, which makes them
particularly interesting to study. Unfortunately, Einstein-Hilbert
gravity in two dimensions is too simple - the action functional is
just the Gauss-Bonnet term yielding the Euler number of the space-time
manifold. The simplest way to generalize the Hilbert action by
introducing an additional scalar field -- the dilaton -- leads to
Generalized Dilaton Theories described by the action \eqref{eq:GDT},
which was first introduced in full generality in
\cite{Odintsov:1991qu,Banks:1991mk,Russo:1992yg}.\footnote{For a
  concise history of the topic cf.  e.g. the review
  \cite{Grumiller:2002nm}.} In their first order formulation described
by the action \eqref{eq:FOG}, which was first given in this general
form in \cite{Schaller:1994es,Strobl:1994yk}, the dynamics of this
class of theories indeed simplifies such that all classical solutions
can be obtained explicitly \cite{Kummer:1995qv} and a global
classification of all generalized dilaton theories can be given
\cite{Klosch:1996qv}. In this context a background independent
\cite{Grumiller:2003sk} and nonperturbative path integral quantization
of first order gravity without matter \cite{Kummer:1996hy}, coupled to
real scalar fields \cite{Kummer:1997jj,Kummer:1998zs,Grumiller:2001ea},
of dilaton supergravity in two dimensions
\cite{Bergamin:2004us,Bergamin:2004aw} and of dilaton gravity in Euclidean space-time
\cite{Bergamin:2004} is possible. Also the role of time was discussed
\cite{Schaller:1992np} in the context of Dirac quantization of dilaton
gravity
\cite{Cangemi:1995yz,Louis-Martinez:1993eh,Kuchar:1994zk,Kuchar:1996zm}\footnote{These
  are only the most influential papers on that topic. A complete list of references
  can be found in chapter 9 of \cite{Grumiller:2002nm}.}.

In this thesis I consider the exact path integral quantization of
first order gravity in two dimensions coupled to Dirac
fermions.\footnote{Such models were first considered on the level of
  classical solutions \cite{Kummer:1992ef,Obukhov:1994wv,Cavaglia:1998yp} and used later on in
  studies of the evaporation of charged CGHS \cite{Callan:1992rs}
  black holes \cite{Nojiri:1992st,Ori:2001xc}.} Interesting by itself
because it has been a blind spot in the literature on quantum dilaton
gravity until now, the main motivation of this analysis is the
comparison to the scalar case
\cite{Kummer:1997jj,Kummer:1998zs,Grumiller:2001ea}. The quantum
equivalence between the massive Thirring model and the Sine-Gordon
model in flat Minkowski space \cite{Coleman:1975pw,Coleman:1974bu} has
recently been applied \cite{Frolov:2005ps,Frolov:2006is,Thorlacius:2006tf} to
semi-classical studies of two-dimensional dilaton gravity
electrodynamics, with the matter model treated as a quantized theory
on a classical background described by the dilaton gravity theory. As
it is already known how to nonperturbatively quantize dilaton gravity
coupled to scalar fields in the path integral formalism, the study of
the corresponding fermionic systems is a prerequisite for
investigating whether and how bosonization could carry over to the
quantum (dilaton) gravity regime.

The structure of this thesis is as follows: 
\blist

\item Chapter~\ref{ch:classical} presents the constraint analysis of
  two-dimensional dilaton gravity in its first order formulation
	coupled to Dirac fermions. After calculating the algebra of first
  class constraints the BRST charge is constructed, the phase space is
  enlarged by a ghost sector and Eddington-Finkelstein gauge is fixed by constructing a BRST invariant Hamiltonian.

\item In chapter~\ref{ch:nonpert} the first order gravity sector of
  the theory is quantized nonperturbatively using the Feynman path
  integral approach, resulting in an effective action which still
  depends on the Dirac fermions.

\item The remaining path integral quantization of the Dirac fermions
  is carried out by means of perturbation theory in
  chapter~\ref{ch:matter}. For massless and minimally coupled fermions
  chiral and conformal anomalies are used to calculate the one-loop
  approximation of the total effective action of the system. The
  gravitationally induced four-fermi scattering vertices are derived.

\item Chapter~\ref{ch:conclusions} contains a summary of the results, conclusions and an outlook to possible further developments.

\elist


\chapter{Classical Analysis}\label{ch:classical}

\section{Dilaton Gravity in Two Dimensions with Fermions}

The field theoretic system considered in this thesis consists of two
sectors, a geometric one and a matter sector. The geometric sector
consists of a scalar field $X$ called the dilaton field and the metric
$g_{\mu\nu}$ on a (1+1) dimensional space-time manifold with signature
$(+-)$ and is described by the Generalized Dilaton Gravity action
\beq
\label{eq:GDT}
\act = \frac{1}{2} \int \extd^{2}x \, \sqrt{-g}\; \Big[ X R  -  U(X)\; (\nabla X)^{2} + 2V(X) \Big] \, ,
\eeq
with $R$ being the Ricci scalar associated with the metric
$g_{\mu\nu}$ and $\nabla$ the Levi-Civita connection. The functions
$U(X)$ and $V(X)$ are specifying the model under consideration. Table
\ref{tab:1} gives a list of models fitting into this action.

\begin{table}[t]
\centering\hspace*{-2.5truecm}
\fbox{
\begin{tabular}{|l||>{$}c<{$}|>{$}c<{$}||>{$}c<{$}|} 
\hline
Model (cf.~\eqref{eq:GDT} or \eqref{eq:FOG})& U(X) & V(X) & w(X) $\,\,(cf.~\eqref{eq:w})$ \\ \hline \hline
1.~Schwarzschild  \cite{Thomi:1984na} 
& -\frac{1}{2X} & -\lambda^2 & -2\la^2\sqrt{X} \\
2.~Jackiw-Teitelboim \cite{Teitelboim:1983ux,Jackiw:1985je} & 0 & -\Lambda X & -\frac12 \Lambda X^2 \\ 
3.~Witten BH/CGHS \cite{Witten:1991yr,Callan:1992rs} & -\frac{1}{X} & -2b^2 X & -2b^2X \\
4.~CT Witten BH \cite{Witten:1991yr,Callan:1992rs} & 0 & -2b^2  & -2b^2X \\
5.~SRG ($D>3$) & -\frac{D-3}{(D-2)X} & -\lambda^2 X^{(D-4)/(D-2)} & -\la^2\frac{D-2}{D-3} X^{(D-3)/(D-2)}\\  
6.~$(A)dS_2$ ground state \cite{Lemos:1994py} &  -\frac{a}{X} & -\frac{B}{2}X  & a\neq 2:\,\,-\frac{B}{2(2-a)} X^{2-a}\\
7.~Rindler ground state \cite{Fabbri:1996bz} & -\frac{a}{X} & -\frac{B}{2} X^a  & -\frac{B}{2} X \\
8.~BH attractor \cite{Grumiller:2003hq} & 0 & -\frac{B}{2}X^{-1} & -\frac{B}{2}\ln{X} \\ \hline
9.~All above: $ab$-family \cite{Katanaev:1997ni} & -\frac{a}{X} & -\frac{B}{2} X^{a+b} & b\neq-1:\,\,-\frac{B}{2(b+1)}X^{b+1} \\   \hline 
10.~Liouville gravity \cite{Nakayama:2004vk} & a & b e^{\al X} & a\neq-\al:\,\,\frac{b}{a+\al}e^{(a+\al)X} \\
11.~Scattering trivial \cite{Grumiller:2002dm} & $generic$  & 0 & $const.$ \\
12.~Reissner-Nordstr\"om \cite{Reissner:1916} & -\frac{1}{2X} & -\lambda^2 + \frac{Q^2}{X} & -2\la^2\sqrt{X}-2Q^2/\sqrt{X}\\
13.~Schwarzschild-$(A)dS$ \cite{Hawking:1982dh} & -\frac{1}{2X} & -\lambda^2 - \ell X & -2\la^2\sqrt{X} - \frac23 \ell X^{3/2} \\
14.~Katanaev-Volovich \cite{Katanaev:1986wk} & \alpha & \beta X^2 - \Lambda  & \int^X e^{\al y}(\be y^2-\La)\extd y\\
15.~Achucarro-Ortiz \cite{Achucarro:1993fd} & 0 &  \frac{Q^2}{X} - \frac{J}{4X^3} - \Lambda X &  Q^2\ln{X} + \frac{J}{8X^2} - \frac12 \La X^2 \\
16.~KK reduced CS \cite{Guralnik:2003we,Grumiller:2003ad} & 0 & \frac12 X(c-X^2) & -\frac18 (c-X^2)^2 \\ 
17.~Symmetric kink \cite{Bergamin:2005au} & {\rm generic} & -X\Pi_{i=1}^n(X^2-X_i^2) & $cf.~\cite{Bergamin:2005au}$ \\
18.~2D type 0A/0B \cite{Douglas:2003up,Gukov:2003yp} & -\frac{1}{X} & -2b^2X+\frac{b^2q^2}{8\pi}  & -2b^2X+\frac{b^2q^2}{8\pi}\ln{X}\\
19.~exact string BH \cite{Dijkgraaf:1992ba,Grumiller:2005sq} & $cf.~\cite{Grumiller:2005sq}$ & $cf.~\cite{Grumiller:2005sq}$  & $cf.~\cite{Grumiller:2005sq}$ \\
\hline
\end{tabular}
}
\caption{List of models (taken from \cite{Grumiller:2006rc})}
\label{tab:1}
\end{table}

Introducing a Zweibein $e_a^\mu$ and a dual dyad 1-form $e^a = e^a_\mu
\extd x^\mu$ which are related to each other and to the metric
by\footnote{For conventions used throughout this thesis, see App.
  \ref{conventions}.}
\begin{align}\label{eq:metriczweibein}
g_{\mu\nu} = e^a_\mu \eta_{ab} e^b_\nu\,, && e^a_\mu e^\mu_b = \de^a_b\,, && e^\mu_a e^a_\nu = \de^\mu_\nu\,,
\end{align}
the \Ix{first order gravity action} 
\beqa\label{eq:FOG}
S^{\rm FOG} & = & \int_{\mathcal{M}_2} \left[X_aT^a+X \cR +\epsilon\mathcal{V} (X^aX_a,X)\right]
\eeqa
is equivalent to \eqref{eq:GDT} for potentials \cite{Grumiller:2002nm}
\eq{{\cal V}(X^aX_a,X) = \frac{X^aX_a}{2} U(X) + V(X)\,.}{eq:dilpot}
The quantity $\epsilon$ denotes the volume form on the manifold.  The
fields $X^a$ are Lagrange multipliers for the torsion 2-form, which in
turn is defined by Cartan's first structure equation
\beq\label{eq:cartan1}
T^a = {D^a}_b e^b = ({\de^a}_b \, \extd + {\epsilon^a}_b \, \om) \wedge e^b\,,
\eeq
with ${D^a}_b = {\de^a}_b \, \extd + {\epsilon^a}_b\, \om$ being the
local Lorentz covariant derivative written as a 1-form. The spin
connection is required to be metric compatible and thus antisymmetric
\cite{nakaharageometry}, ${\om^a}_b = - {\om_b}^a$, but first order
gravity allows for torsion if $U(X)\neq 0$. Because the Lorentz group
in two dimensions, $SO(1,1)$, has only one generator,
${\epsilon^a}_b$, the spin connection can then be expressed as
${\om^a}_b = {\epsilon^a}_b\, \om$. The quantity $\cR$ in
\eqref{eq:FOG} is the curvature 2-form related to the spin connection
by Cartan's second structure equation
\beqa\label{eq:cartan2}
{\cR^a}_b & = & {D^a}_c \, {\om^c}_b\ = {\epsilon^a}_b\, \extd \om + {\epsilon^a}_c {\epsilon^c}_b \, \om \wedge \om = {\epsilon^a}_b \extd \om \\\label{curv2form}
\cR       & := & \frac{1}{2} {\epsilon^a}_b {\cR^b}_a = \extd \om\,.
\eeqa
It is related to the Ricci scalar by $R = 2 \ast \extd \om$. Using
\eqref{eq:starquadrat} and \eqref{eq:starepsilon} this relation is
inverted, yielding
\eq{\extd \om = \frac{R}{2}\epsilon\,.}{eq:domegaR}
This Ricci scalar however differs from the one in eq.~\eqref{eq:GDT}
by the torsion part of the spin connection. Only for theories with
vanishing torsion $U(X)=0$ both definitions coincide (also cf. the
remarks in sec.~\ref{ssec:effgeom} on eqs.~\eqref{eq:REFsol} and
\eqref{eq:Rtildeas}).

\subsection{Equivalence between First and Second Order Formulation}

The equivalence between first order gravity \eqref{eq:FOG} and
generalized dilaton theories \eqref{eq:GDT} is established in two
steps \cite{Grumiller:2002nm}. First one decomposes the spin
connection into a Levi-Civita part $\tilde{\om}$ and a torsion part
$\om_{\rm T}$,
\eq{\om = \tilde{\om} + \om_{\rm T}\,.}{eq:composespincon}
The Levi-Civita spin connection explicitly reads
\eq{\tilde{\om} = e_a (\ast \extd e^a)\,,}{eq:LCconnection}
has vanishing torsion $\tilde{T}^a = (\tilde{D}e)^a = 0$ and will be
considered fixed from now on. The second step is to use the equations
of motion following from varying eq.~\eqref{eq:FOG} with respect to
$X^a$ and $\om_{\rm T}$,
\beqa\label{eomXa}
0 & = & \underbrace{(\tilde{D}e)^a}\limits_{=0} + {\epsilon^a}_b \,\om_{\rm T}\wedge e^b + \epsilon \pd{\cV}{X_a}\\\label{eomtildeom}
0 & = & \extd X + X^a \epsilon_{ab} e^b\,,
\eeqa
to eliminate both the Lagrange multipliers $X^a$ and the torsion part $\om_{\rm T}$
of the spin connection. Wedge multiplying
eq.~\eqref{eomtildeom} with $e^c$ from the right, using
eq.~\eqref{volform1} and taking the Hodge dual yields the solution for
the Lagrange multipliers
\eq{X^a = \ast (e^a \wedge \extd X)\,.}{eq:solXa}
Decomposing $\om_{\rm T} = \om_{Ta}e^a$, a similar calculation for the class
of potentials \eqref{eq:dilpot} yields from eq.~\eqref{eomXa}
\eq{\om_{\rm T} = U(X)X_a e^a = U(X)\ast (e_a \wedge \extd X)e^a\,.}{eq:solomT}
The torsion part of the spin connection thus vanishes if and only if
$U(X)=0$. Inserting eq.~\eqref{eq:composespincon} into \eqref{eq:FOG}
gives after using the torsionlessness of the Levi-Civita connection,
eq.~\eqref{eq:domegaR} and a partial integration
\eq{S^{\rm FOG}_1 = \int\limits_{\cM_2} \Big[ \epsilon \frac{\tilde{R}}{2} X +
  \epsilon V(X) - \extd X\wedge \om_{\rm T} + X^a\, \epsilon_{ab}\, \om_{\rm T}
  \wedge e^b + \epsilon \frac{U(X)}{2} X^a X_a \Big]\,.}{eq:SFOG1}
After inserting eq.~\eqref{eq:solomT} into eq.~\eqref{eq:SFOG1} and
using
\beqs
X^a X_a = - g^{\mu\nu} (\nabla_\mu X) (\nabla_\nu X)\,,
\eeqs
the third and fourth term in \eqref{eq:SFOG1} cancel and the last one
yields the kinetic term for the dilaton in \eqref{eq:GDT}.  This
proves the equivalence of the first order gravity action
eq.~\eqref{eq:FOG} and the dilaton gravity action eq.~\eqref{eq:GDT}
for potentials of the form eq.~\eqref{eq:dilpot}. The above
calculation also holds for first order gravity coupled to matter
fields which do neither couple to the Lagrange multipliers $X^a$ nor
to the spin connection $\om$, so that the equations of motion
\eqref{eomXa} and \eqref{eomtildeom} do not receive additional matter
contributions. It holds in particular for fermions which in general do
couple to the spin connection through their kinetic term, but not in
dimension two (cf.  sec.~\ref{ssec:fermions}). Under the same
restrictions the equivalence proof given in \cite{Kummer:1996hy} for
the corresponding quantum theories without matter also applies to the
case with matter.

\subsection{Classical Solutions}\label{sec:classsol}

First order gravity \eqref{eq:FOG} is an integrable system
\cite{Grumiller:2002nm}, i.e. all classical solutions can be obtained
in a closed form. Although integrability is lost in general if matter
is added, only one integration cannot be carried out explicitly. First
denote the variations of the matter Lagrangian by
\beq
W^\pm := \frac{\de S^{\rm m}}{\de e^\mp}\,,\quad W:=\frac{\de S^{\rm m}}{\de X}\,,
\eeq
where the indices $\pm$ denote light cone coordinates
\eqref{eq:lccoord} in tangent and cotangent space. A coupling of
matter to the spin connection is not considered, because neither
fermions in two dimensions nor scalars do couple in such a way. The
equations of motion corresponding to variations $\de \om$, $\de
e^\mp$, $\de X$ and $\de X^\mp$ respectively are
\beqa\label{eomom}
0 & = & \extd X + X^- e^+ - X^+ e^- \\\label{eome}
0 & = & (\extd \pm \om) X^\pm \mp \cV e^\pm + W^\pm\\\label{eomX}
0 & = & \extd \om + \epsilon \pd{\cV}{X} + W \\\label{eomXmp}
0 & = & (\extd \pm \om) e^\pm + \epsilon \pd{\cV}{X^\mp}\,.
\eeqa

In a patch with $X^+ \neq 0$ a new one-form $Z=e^+/X^+$ can be
introduced and eq.~\eqref{eome} with the upper sign gives
\beqa 
\om = -\frac{\extd X^+}{X^+} + Z \cV - \frac{W^+}{X^+} \,,
\eeqa
while eq.~\eqref{eomom} yields
\beqa
e^- = \frac{\extd X}{X^+} + X^- Z\,.
\eeqa
Using these two solutions plus eq.~\eqref{eomXmp} with the upper sign
(and $\epsilon = e^+ \wedge e^- = Z\wedge \extd X$) one derives the
relation
\beqa
\extd Z - U(X)\extd X \wedge Z = \frac{W^+}{X^+} \wedge Z\,.
\eeqa
If $W^+ = 0$, which corresponds to (anti)chiral fermions (cf.
eq.~\eqref{eq:Lkin}) and (anti)self-dual scalars\footnote{The
  (anti)self-dual components of a real scalar field $\phi$ are defined
  by $\phi^\pm = \ast (\extd \phi \wedge e^\pm)$. In these terms
  the action for a non-minimally coupled free scalar reads $S_\phi =
  \int \extd^2 x \sqrt{-g} f(X) \phi^+ \phi^-$.}, the Ansatz $Z =
\hat{Z} e^{Q(X)}$ (with $Q(X)$ defined in \eqref{eq:wI} below) leads to $\extd
\hat{Z}=0$, which can be integrated by use of the Poincaré Lemma to
$\hat{Z} = \extd f$.  This exact integration is not possible for general
matter contributions. The classical solutions obtained in this way are
\beqa\label{eq:sole+}
e^+ & = & X^+ e^{Q(X)} \extd f \\\label{eq:sole-}
e^- & = & \frac{\extd X}{X^+} + X^- e^{Q(X)} \extd f \\\label{eq:solom}
\om & = & -\frac{\extd X^+}{X^+} + e^{Q(X)} \cV \extd f\,.
\eeqa
The theory exhibits an absolute, i.e. in space and time, conserved
quantity \cite{Mann:1992yv,Frolov:1992xx,Kummer:1995qv} (with $w(X)$ defined
in \eqref{eq:w} below)
\beqa
\extd \cC & = & 0 \\\label{eq:conservedquantity}
\cC & = & \cC^g + \cC^m  \\
    & = & e^{Q(X)} X^+ X^- + w(X) + \cC^m  = \cC_0 = {\rm const.}\\
\extd \cC^m & = & e^{Q(X)} (X^+W^- + X^- W^+)\,.
\eeqa
The line element for the solutions reads
\eq{(\extd s)^2 = \eta_{ab}e^a \otimes e^b = 2 e^{Q(X)} \extd f
  \otimes \big[ \extd X + X^+X^- e^{Q(X)}\extd f\big]\,.}{eq:lineel}
The quantity $X^+X^-e^{Q(X)}$ encodes information about the horizons of
the solutions, which lie at $X^+X^- = 0$.  In the absence of matter
$W^\pm = W = 0$ it can with the help of the conserved quantity be
rewritten as $\cC_0 - w(X)$, which immediately shows the existence of
a Killing vector $\partial/\partial f$ for all dilaton gravity models.
The corresponding Killing norm reads $e^{Q(X)}(\cC_0 - w(X))$.

For the sake of completeness it should be mentioned that in addition
to the vacuum solutions obtained above which are labeled by the
conserved quantity $\cC^{g}$, isolated solutions (cf. e.g.
\cite{Bergamin:2005au}) with constant dilaton $X=X_{\rm CDV}$,
so-called constant dilaton vacua (CDV), exist if $X^\pm$ both vanish
in some open region. Eq.~\eqref{eome} restricts the value of the
dilaton to be one of the zeroes of the potential $V(X)$, $V(X_{\rm
  CDV}) = 0$. The curvature for these solutions is constant (cf.
eq.~\eqref{eomX}),
\eq{\extd \om = - \epsilon V'(X_{\rm CDV})\,,\quad R = 2\ast \extd \om
  = -2 V'(X_{\rm CDV})\,.}{eq:CDVcurv}
The geometry of the constant dilaton vacua is thus Anti-de Sitter space ($R>0$),
de Sitter space ($R<0$) and Minkowski or Rindler space ($R=0$).\footnote{In
  the conventions of \cite{Grumiller:2002nm}, which are adopted in
  this thesis, the sphere has negative curvature.}

\subsection{Conformally Related Theories}\label{ssec:conf}

Although first order gravity \eqref{eq:FOG} is not conformally
invariant, dilaton dependent conformal transformations
\begin{align}\label{eq:FOGconftrans}
X^a \mapsto \frac{X^a}{\Om}\,, && e^a \mapsto e^a \Om\,, && \om \mapsto \om + X_a e^a \frac{\extd \ln \Om}{\extd X}
\end{align}
with a conformal factor $\Om = e^{\frac{1}{2}\int\limits^X (U(y) -
  \tilde{U}(y)) \extd y}$ map a model with potentials $(U(X),V(X))$ to
one with\footnote{The action picks up a boundary term
  $\frac{1}{2}\int\limits_{\cM_2} \extd (X X_a
  e^a(U(X)-\tilde{U}(X)))$.}  $(\tilde{U}(X),\tilde{V}(X)= \Om^2
V(X))$. Thus one can always transform to a conformal frame with
$\tilde{U}=0$, which simplifies the equations of motion considerably,
do calculations there and afterwards transform back to the original
conformal frame. The expression
\eq{w(X) = \int\limits^X e^{Q(y)} V(y) \extd y}{eq:w}
is invariant under conformal transformations, whereas
\eq{Q(X) = \int\limits^X U(y) \extd y}{eq:wI}
captures the information about the conformal frame.

\subsection{Fermions}\label{ssec:fermions}

The action for a Dirac fermion in two dimensions (see
App.~\ref{conventions} for conventions) consists of a kinetic
term\footnote{The derivative acting on both sides is defined as
  $A\lrpd{}B = A
  \partial B - (\partial A) B$. The sign difference between
  \cite{Meyer:2005fz} and this thesis for $S^{(\rm kin)}$ stems from
  the differing sign of $\epsilon^{ab}$.}
\beq\label{eq:fermkin}
S^{(\rm kin)}  =  \frac{i}{2} \int\limits_{{\cal M}_2} f(X)\;(*e^a) \wedge (\chib \ga_a \overleftrightarrow{\extd}\chi)\,,
\eeq
and a general self-interaction
\beq
S^{(\rm SI)}   =  - \int\limits_{{\cal M}_2} \epsilon h(X) g(\chib\chi)\label{eq:fermSI}\,.
\eeq
The functions $f(X)$ and $h(X)$ are the dilaton coupling functions. If
they are constant one speaks of minimally coupled matter, and of
non-minimally coupled matter otherwise. Because of the Grassmann
property of the spinor field the self-interaction can in two
dimensions at most include a mass term and a Thirring term,
\eq{g(\chib\chi) = m\chib\chi + \la(\chib\chi)^2\,.}{eq:SITaylor}
The constant contribution in this Taylor expansion of $g$ has been set
to zero because it would only shift the dilaton potential $V(X)$.

In two dimensions the kinetic term is independent of the spin
connection. In arbitrary dimension, the kinetic term for fermions on a
curved background reads\footnote{The abbreviation ``h.c.'' means
  hermitian conjugate.}  \cite{Birrell:1989}
\beqs
\frac{i}{2} \int d^n x \det(e^a_\mu) \left[ e_a^\mu (\chib \ga^a (\partial_\mu + {\om_\mu}^{bc}\Si_{bc})\chi) + \mathrm{h.c.}\right]\,.\eeqs
For $n=2$ there is only one independent generator of Lorentz
transformations $\Si_{01} = \frac{1}{4}[\ga_0,\ga_1]=\frac{\ga_*}{2}$,
and with $\{\ga_a,\ga_*\}=0$ the terms in \eqref{eq:fermkin}
containing the spin connection vanish,
\beqas
&   & \frac{i}{4} (\ast e^a) \wedge \om \chi^\dagger(\ga^0\ga_a\ga_* - \ga_*\ga_a^\dagger \ga^0)\chi\\
& = & \frac{i}{4} (\ast e^a) \wedge \om \chi^\dagger\underbrace{(\ga^0\ga_a - \ga_a^\dagger \ga^0)}\limits_{= 0} \ga_*\chi = 0\,.
\eeqas
This simplifies the constraint algebra and is one of the reasons why
the path integral over the geometric sector can be carried out
explicitly. In higher dimensions there would be a multiplicative
coupling between the Vielbeine and the spin connection. Only in two
dimensions the whole Lagrangian turns out to be at most linear in the
geometric fields $e^a$ and $\om$.

\section{Hamiltonian Analysis}\label{sec:analysis}

Henceforth we denote the canonical coordinates and momenta by
\beqa\label{qbari}
\ol{q}^i & = & (\om_0,e_0^-,e_0^+)\\\label{qi}
q^i      & = & (\om_1,e_1^-,e_1^+),\hspace{9.5mm} i=1,2,3 \\\label{pi}
p_i      & = & (X,X^+,X^-)\\\label{Qj}
Q^\al      & = & (\chi_0,\chi_1,\chid_0,\chid_1),\hspace{3mm} \al=0,1,2,3\,.
\eeqa
The canonical structure on the phase space is given by graded
equal-time Poisson brackets
\beqa\nonumber
\spoiss{q^i(x)}{p_j(y)} & = & \de^i_j \de(x-y) \\\label{poisson}
\spoiss{Q^\al(x)}{P_\be(y)} & = & - \de^\al_\be \de(x-y)\,,
\eeqa
where the $P_\be$ are canonical momenta for the spinors defined by
$P_\be := \pld{\cL}{Q^\be}$ ($\partial^L$ is the usual left derivative
on Grassmann numbers). The graded Poisson bracket is defined as
\cite{Henneaux:1992}
\beqa\nonumber
\spoiss{F}{G} & = & \int dz \left[ \left( \pd{F}{q^i(z)} \pd{G}{p_i(z)} - \pd{F}{p_i(z)} \pd{G}{q^i(z)} \right)\right. \\\label{eq:gPB}
              &   & \left. + (-)^{\eps(F)} \left( \pld{F}{Q^\al(z)} \pld{G}{P_\al(z)} - \pld{F}{P_\al(z)} \pld{G}{Q^\al(z)} \right)\right]\,,
\eeqa
with $(q^i,p_i)$ and $(Q^\al,P_\al)$ being bosonic ($\eps(q^i)=
\eps(p_i)=0$) and fermionic ($\eps(Q^\al)=\eps(P_\al)=1$) canonical
pairs, respectively. $F$ and $G$ are functions on the phase space with
definite Grassmann parities $\eps(F)$, $\eps(G)$. In this chapter
three main properties of this bracket will be frequently used, namely
the symmetry property, the Leibniz rule and the graded Jacobi identity
\beqa\label{eq:gPBsymmetry}
\spoiss{F}{G} & = & (-)^{\eps(F)\eps(G) + 1} \spoiss{G}{F} \\\label{eq:gPBLeibnitz}
\spoiss{F}{G_1 G_2} & = & \spoiss{F}{G_1} G_2 + (-)^{\eps(F)\eps(G_1)} G_1 \spoiss{F}{G_2}\\\nonumber
0 & = & \hspace{3.2mm}(-)^{\eps(F_1)\eps(F_3)}\poiss{\poiss{F_1}{F_2}}{F_3} \\\nonumber
  & & + (-)^{\eps(F_1)\eps(F_2)}\poiss{\poiss{F_2}{F_3}}{F_1} \\\label{eq:gPBJacobi}
  & & + (-)^{\eps(F_2)\eps(F_3)}\poiss{\poiss{F_3}{F_1}}{F_2} \,.
\eeqa
Henceforth the graded Poisson bracket will just be referred to as
``Poisson bracket'' except for where the distinction is crucial.

\subsection{Primary and Secondary Constraints}

The system under consideration admits both primary first and second
class constraints. A look at the component form of the Lagrangian,
\beqa\label{eq:Ltotal}
\cL                 & = & {\cal L}^{\rm FOG} + {\cal L}^{\rm (kin)} + {\cal L}^{\rm (SI)} \\\nonumber
{\cal L}^{\rm FOG}  & = & \tilde{\epsilon}^{\nu \mu}(X^+(\partial_\mu - \omega_\mu)e^-_\nu + X^-(\partial_\mu + \omega_\mu)e^+_\nu \\\label{eq:LFOG}
                &   & + X \partial_\mu \omega_\nu) + (e){\cal V}(X^+X^-;X) \\\nonumber
{\cal L}^{\rm (kin)}  & = & \frac{i}{\sqrt{2}} f(X)\left[ -e_0^+(\chid_0\plr{1}\chi_0)
                                                     +e_0^-(\chid_1\plr{1}\chi_1)\right.\\\label{eq:Lkin}
           & & \hspace{19mm}\left.                   +e_1^+(\chid_0\plr{0}\chi_0)
                                                     -e_1^-(\chid_1\plr{0}\chi_1)
                                               \right]\\\label{eq:LSI}
{\cal L}^{\rm (SI)}   & = & -(e) h(X) g(\chid_1 \chi_0 + \chid_0 \chi_1)\,,
\eeqa
where $(e)= \det e_\mu^a$, shows that there do not occur any time
(i.e. $x^0$) derivatives of the $\ol{q}^i$ and thus the corresponding
canonical momenta $\ol{p}_i:=\pd{\cL}{\ol{q}^i}$ are constrained to
zero,
\eq{\ol{p}_i \approx 0\,.}{eq:pbarizero}
The symbol $\approx$ denotes weak equality, i.e. equality on the
constraint surface.

Because the kinetic term for the fermions is of first order in time
derivatives, the fermion momenta $P_\al$ are related to the spinor
components by another set of primary constraints,
\beqa\label{Phi0}
\Phi_0 & = & P_0 + \frac{i}{\sqrt{2}}f(p_1) q^3 Q^2 \approx 0\\\label{Phi1}
\Phi_1 & = & P_1 - \frac{i}{\sqrt{2}}f(p_1) q^2 Q^3 \approx 0\\\label{Phi2}
\Phi_2 & = & P_2 + \frac{i}{\sqrt{2}}f(p_1) q^3 Q^0 \approx 0\\\label{Phi3}
\Phi_3 & = & P_3 - \frac{i}{\sqrt{2}}f(p_1) q^2 Q^1 \approx 0\,,
\eeqa
which have nonvanishing Poisson brackets with each other,
\beqa\label{Calphabeta}\nonumber
C_{\al\be}(x,y) & := & \spoiss{\Phi_\al(x)}{\Phi_\be(y)} \\
                & = & i\sqrt{2}f(X)
                      \left(
                      \begin{matrix}
                      0      & 0     & -e_1^+ & 0     \\
                      0      & 0     & 0      & e_1^- \\
                      -e_1^+ & 0     & 0      & 0     \\
                      0      & e_1^- & 0      & 0
                      \end{matrix}
                      \right)\de(x-y)\,,
\eeqa
and thus are of second class according to Dirac's classification of
constraints \cite{Dirac:1996} if the matrix \eqref{Calphabeta} has
maximal rank. In Eddington-Finkelstein gauge $(e_0^-,e_0^+)=(1,0)$, which will become very important later on, the condition for a horizon is $e_1^-=0$.  The Dirac matrix
\eqref{Calphabeta} thus has maximal rank away from horizons and the
four constraints \eqref{Phi0}-\eqref{Phi3} are all of second class.
On horizons it has rank two, but the rank never vanishes for a
nondegenerate metric. Thus on a horizon second class constraints are
converted into first class ones, i.e. on the horizon more gauge
symmetries are present than away from it. This phenomenom was first
mentioned by t'Hooft \cite{'tHooft:2004ek} and appears in several
settings, see \cite{Bergamin:2005pg} and references therein. As
mentioned in sec.~\ref{sec:extens}, it also seems to be connected to
the problem of black hole universality.

The $\Phi_\al$ are independent of the $\ol{q}^i$ and thus
commute even in the strong sense with the other three primary
constraints $\ol{p}_i$,
\eq{\spoiss{\ol{p}_i}{\Phi_\al} = 0\,.}{eq:gPBpbariPhial}

Having computed all the momenta, one obtains the Hamiltonian density
from the Lagrangian \eqref{eq:Ltotal} by means of ordinary Legendre
transformation,
\beqa\nonumber
\cal H         & =  & \dot{Q}^\al P_\al + p_i \dot{q}^i - \cal L \\\label{wholehamiltonian}
               & =: & {\cal H}_{FOG} + {\cal H}_{kin} + {\cal H}_{SI} \\\nonumber
{\cal H}_{FOG} & = & X^+ (\partial_1 - \om_1)e_0^- + X^-(\partial_1 + \om_1)e_0^+ + X\partial_1 \om_0 - (e){\cal V}\\\label{eq:HFOG}
               &   & +(X^+e_1^- - X^-e_1^+)\om_0\\\label{eq:Hkin}
{\cal H}_{kin} & = & \frac{i}{\sqrt{2}}f(X)\left[e_0^+(\chid_0 \lrpd{1} \chi_0) - e_0^-(\chid_1 \lrpd{1} \chi_1) \right]\\\label{eq:HSI}
{\cal H}_{SI}  & = & (e)h(X)g(\chib\chi)
\eeqa

To deal with the second class constraints one introduces the Dirac
bracket \cite{Dirac:1996,Henneaux:1992}
\beq\label{diracbracket}
\dirac{F_x}{G_y} := \spoiss{F_x}{G_y} - \int dzdw \; \spoiss{F_x}{\Phi_\al(z)} C^{\al\be}(z,w) \spoiss{\Phi_\be(w)}{G_y}\,,
\eeq
with the subscripts $x$, $y$ denoting functions evaluated at the
corresponding points in space. It inherits the properties
\eqref{eq:gPBsymmetry}, \eqref{eq:gPBLeibnitz} and
\eqref{eq:gPBJacobi} from the graded Poisson bracket.
$C^{\al\be}(x,y)$ is the inverse of the matrix-valued distribution
\eqref{Calphabeta} defined such that
$$\int dy \left(\int dx \varphi(x) C_{\al\ga}(x,y)\right)\left(\int dz
  \psi(z) C^{\ga\be}(y,z)\right) = \delta_\al^\be \int dx \varphi(x)
\psi(x)$$ for all test functions $\varphi,\psi$. It reads explicitly
\beq\label{eq:Calbeinv}
C^{\al\be}(x,y) = \frac{i}{\sqrt{2}f(X)}\, \left(
                      \begin{matrix}
                      0      & 0     & \frac{1}{e_1^+} & 0     \\
                      0      & 0     & 0      & -\frac{1}{e_1^-} \\
                      \frac{1}{e_1^+} & 0     & 0      & 0     \\
                      0      & -\frac{1}{e_1^-} & 0      & 0
                      \end{matrix}
                      \right)\de(x-y)\,.
\eeq

Demanding the primary first class constraints $\ol{p}_i$ not to change
during time evolution, i.e.\footnote{Henceforth, a prime in a Dirac or
  Poisson bracket means evaluation of the function at a point $x'$,
  whereas a prime on a function not being an argument of some bracket
  denotes differentiation with respect to the argument of the
  function.}
$$G_i := \dot{\ol{p}}_i = \dirac{\ol{p}_i}{{\cal H}'} = \spoiss{\ol{p}_i}{{\cal
    H}'} \approx 0\,,$$
yields secondary constraints
\beqa                                                                                         \label{G1}
G_1 & = & G_1^g                                                                             \\\label{G2}
G_2 & = & G_2^g + \frac{i}{\sqrt{2}}f(X)(\chid_1 \lrpd{1} \chi_1) + e_1^+ h(X) g(\chib\chi) \\\label{G3}
G_3 & = & G_3^g - \frac{i}{\sqrt{2}}f(X)(\chid_0 \lrpd{1} \chi_0) - e_1^- h(X) g(\chib\chi)
\eeqa
with $G_i^g$ being the first order gravity constraints \cite{Kummer:1996hy}
\beqa\label{G1g}
G_1^g & = & \partial_1 X + X^- e_1^+ - X^+ e_1^-                                            \\\label{G2g}
G_2^g & = & \partial_1 X^+ + \om_1 X^+ - e_1^+ \cal V                                       \\\label{G3g}
G_3^g & = & \partial_1 X^- - \om_1 X^- + e_1^- \cal V\,.
\eeqa
The Hamiltonian density turns out to be constrained to zero, as
expected for a generally covariant system%
\cite{Henneaux:1992},
\beq\label{constrainedhamiltonian}
{\cal H} = -\ol{q}^i G_i\,,
\eeq
where a boundary term $\int \partial_1(p_i \ol{q}^i)$ on the right
hand side has been dropped. Working with smeared constraints
($\eta(x^1)$ is some test function)
\eq{\ol{p}_i[\eta] = \int \extd x^1 \eta \ol{p}_i}{eq:pbarismeared}
and with the Hamiltonian rather than the Hamiltonian density one finds
that the secondary constraints \eqref{G1}-\eqref{G3} aquire boundary
terms
\eq{G_i[\eta] = \int \extd x^1 G_i(x^1) \eta(x^1) - p_i \eta\atbdry1\,,}{eq:Giboundary}
where $\partial \cM_1$ denotes the boundaries of the $x^1$-direction
of space-time. The Hamiltonian with the $\ol{q}^i$ playing the role of
smearing functions turns out to be a sum over constraints even in the
presence of boundaries,
\beq
H = \int \extd x^1 \cH = -G_i[\ol{q}^i]\,.
\eeq
This is true for first order gravity supplemented by a Gibbons-Hawking
boundary term, as shown in \cite{Bergamin:2005pg}, as well as without
any additional boundary terms in the action.

The secondary constraints have vanishing Dirac bracket with the
$\ol{p}_i$ because both the $G_i$ and $\Phi_\al$ are independent of
the $\ol{q}^i$. They also trivially commute with the primary second
class constraints, $\dirac{\Phi_\al}{G_j'} = 0$, because of the
definition of the Dirac bracket. For the same reason the $\Phi_\al$ do
not give rise to new secondary constraints.

\subsection{Algebra of Secondary Constraints}

Dirac conjectured \cite{Dirac:1996} that every first class constraint
generates a gauge symmetry. The proof of this conjecture is possible
in a very general setting \cite{Gitman:1990}, but making some
additional assumptions (see paragraph 3.3.2 of \cite{Henneaux:1992})
to rule out ``pathological'' examples simplifies it.  These
assumptions are fulfilled in our case, because 1. every constraint
belongs to a well defined generation; 2.  the Dirac bracket ensures
that the primary second class constraints do not generate new ones
and, as will be seen below, the secondary constraints are of first
class and thus there are no further constraints generated by them; and
3. every primary first class constraint $\ol{p}_i$ generates
exactly one $G_i$.

To show that the system does not admit any ternary constraints it is
sufficient to show that the algebra of secondary constraints closes
weakly, i.e.
\eq{\dirac{G_i}{G_j '} = \left.C_{ij}\right.^k (x)G_k\;\de(x-x') \approx 0\,,}{eq:structfunc}
and thus the secondary constraints are preserved
under the time evolution generated through the Dirac bracket,
\beqs
\dot{G}_i = \dirac{G_i}{\left.{\cal H}\right.' } = - \left.\ol{q}^j\right. ' \dirac{G_i}{G_j'} \approx 0\,.
\eeqs

To calculate all the Dirac brackets, one first needs the Poisson
brackets $\spoiss{\Phi_\al}{G_j}$. They are rather lengthy and thus
listed in Appendix \ref{app:B1}, together with some remarks on the
actual calculation. The resulting algebra of secondary constraints
reads
\beqa\label{GiGi}
\dirac{G_i}{G_i '} & = & 0 \hspace{5mm} i = 1,2,3 \\\label{G1G2}
\dirac{G_1}{G_2 '} & = & - G_2 \,\de \\\label{G1G3}
\dirac{G_1}{G_3 '} & = & G_3 \,\de \\\label{G2G3}
\dirac{G_2}{G_3 '} & = & \left[ - \sum\limits_{i=1}^3 \pd{{\cal V}}{p_i} G_i + \left(g h' - \frac{h}{f} f'g'\cdot (\chib\chi)\right)G_1 \right]\de\,.
\eeqa
It should be noted that the algebra for minimally coupled fermions
with zero mass and without self-interaction was already computed in
\cite{wal01}. The right hand sides vanish weakly, and thus the
secondary constraints do not generate new constraints and the Dirac
procedure stops at this level. The algebra still closes like in the
case of a compact Lie groups, $[\de_A,\de_B]={f^C}_{AB}(x)\de_C$, but
rather with structure functions than with constants. The second term
in \eqref{G2G3} deserves some remarks.  Without it, the algebra would be just the one obtained for first order gravity without matter \cite{Kummer:1996hy}. It vanishes for minimal
coupling, i.e. for $h=f={\rm const.}$ For $h \propto f$ and $g =
m\chib\chi + \la (\chib\chi)^2$ it becomes proportional to
$f'(g-g'\cdot\chib\chi)G_1\de$, thus a mass term does not change the
constraint algebra at all.  Furthermore, it does not contain
derivatives of the matter field as opposed to the case of scalar
matter (see eq.  E.31 in \cite{Grumiller:2001ea}), where the
additional contribution to $\poiss{G_2}{G_3}$ is proportional to
$\frac{f'}{f}{\cal L}_{scalar}$.

It is noteworthy that if the $x^1$-direction has a boundary, in the
matterless theory only the bracket between the diffeomorphism
constraints eq.~\eqref{G2G3} is modified \cite{Grumiller:2006rc},
aquiring a boundary term of the form
\eq{X(U(X)+X U'(X))X^+X^- - (V(X) - X V'(X))}{eq:G2G3bdry}
which vanishes only for $V\propto X$ and $U \propto 1/X$, i.e. for
models with an $(A)dS_2$ ground state (cf. the second, third and sixth
entry in table~\ref{tab:1}).

In contrast to the matterless theory \cite{Grosse:1992vc} the algebra
generated by the $G_i$ and $p_i$ is no classical finite
W-algebra\footnote{A classical finite W-algebra is, according to
  \cite{deBoer:1995nu}, an algebra with generators $W_\al$ and Poisson
  brackets $\poiss{W_\al}{W_\be} = P_{\al\be}(W_\ga)$, where
  $P_{\al\be}$ are certain polynomials in the generators. In the case
  considered here the Dirac bracket replaces the Poisson bracket.}
anymore,
\beqa\label{G1p1}
\dirac{G_1}{p_1'} & = & 0 \\\label{G23p1G1p23}
\dirac{G_{2/3}}{p_1'} & = & \pm p_{2/3}\de = - \dirac{G_1}{p_{2/3}'} \\
\dirac{G_2}{p_2'} & = & \frac{i}{\sqrt{2}} \frac{f}{q^2} (Q^3 \lrpd{1} Q^1))\de + \frac{1}{2} \frac{q^2}{q^3} h\, g'\cdot (\chib\chi)\de \\
\dirac{G_3}{p_3'} & = & -\frac{i}{\sqrt{2}} \frac{f}{q^3} (Q^2 \lrpd{1} Q^0))\de - \frac{1}{2} \frac{q^2}{q^3} h\, g'\cdot (\chib\chi)\de \\
\dirac{G_3}{p_2'} & = & \Big[ \cV - h\, g - \frac{1}{2} h \, g' \cdot (\chib\chi) \Big]\de = - \dirac{G_2}{p_3'}\,.
\eeqa

The constraints $G_1$ and $G_{2/3}$ on-shell generate local Lorentz
transformations and diffeomorphisms, respectively. The right bracket
in eq.~\eqref{G23p1G1p23} shows that the $X^\pm$ transform as local
Lorentz vectors under the action of $G_1$,
\beq
\de_\ga X^\pm = \ga\dirac{{X^\pm}}{G_1'} = \pm \ga X^\pm \de\,.
\eeq
Eqs.~\eqref{G1G2}~and~\eqref{G1G3} show that $G_2$ and $G_3$ also form
a local Lorentz vector with components $G^+ = G_2$, $G^- = G_3$. The
action of
\beq
G_\mu = -e_{\mu a} G^a
\label{eq:diffgen}
\eeq
on the dilaton is
\beqa\nonumber
\de_\xi X(x') & = & \xi^\mu \dirac{X}{G_\mu'} = -\xi^\mu\Big[ e_\mu^+ \dirac{X}{G_3'} + e_\mu^- \dirac{X}{G_2'}\Big]\\\nonumber
              & = & \xi^\mu\Big[e_\mu^- X^+ - e_\mu^+ X^-  \Big]\de \\
              & = & \xi^\mu (\partial_\mu X) \de\,.
\eeqa
In the last step the equation of motion \eqref{eomom} has been used.
This shows that $G_\mu$ on-shell is the generator of diffeomorphisms.

\subsection{Relation to the Conformal Algebra}\label{sec:witt}

As first noted in \cite{Katanaev:1994qf}, certain linear combinations
of the $G_i$ fulfil the classical Virasoro algebra. In that work first
order gravity coupled to scalar matter was considered, but the same
result holds for the case with fermionic matter. New generators
\begin{align}\label{eq:newgen}
G = G_1\,,&& H_{0/1} = q^1 G_1 \mp q^2 G_2 + q^3 G_3
\end{align}
fulfil an algebra (with $\de' = \pd{\de(x-x')}{x'}$)
\beq\label{eq:alg2}
\begin{array}{ll}
	\dirac{G}{G'}  = 0       & \qquad \dirac{H_i}{H_i'} = (H_1 + H_1')\de' \\
	\dirac{G}{H_i '} = -G \de' & \qquad \dirac{H_0}{H_1'} = (H_0 + H_0')\de' \,.
\end{array}
\eeq
The constraints $H_i$ form the well-known conformal algebra. The total
algebra is the semidirect product of the conformal algebra and an
invariant abelian subalgebra generated by the local Lorentz generator
$G$ \cite{Katanaev:2000kc}. The diffeomorphism part of $H_1$ is just
$-G_0$ from eq.~\eqref{eq:diffgen} above. The Dirac brackets needed
for calculating this algebra are listed in App. \ref{app:B2}.
Defining light cone combinations
\eq{H^\pm = \frac{1}{2}(H_0\pm H_1)\,,}{eq:Hpm}
the conformal part of the algebra \eqref{eq:alg2} reads
\begin{align}\label{eq:alg2lc}
\dirac{H^\pm}{{H^\pm}'} = \pm(H^\pm + {H^\pm}')\de' && \dirac{H^+}{{H^-}'} = 0\,.
\end{align}
The Fourier modes $L_k$, $\ol{L}_k$ defined by
\begin{align}
H^+(x) = \int \frac{dk}{2\pi}\, L_k \,e^{ikx}\,, &&
H^-(x) = \int \frac{dk}{2\pi} \,\overline{L}_k \,e^{ikx}
\end{align}
then obey an algebra
\beqa\nonumber
\dirac{L_k}{L_m} &=& i(m-k) L_{k+m} \\\label{witt}
\dirac{\ol{L}_k}{\ol{L}_m} &=& -i(m-k) \ol{L}_{m+k} \\\nonumber
\dirac{L_k}{\ol{L}_m} &=& 0
\eeqa
which after redefinition $\ol{L} \mapsto - \ol{L}$ becomes the
classical Virasoro algebra. Canonical quantization, i.e.
representing the Virasoro generators as operators and replacing
the Dirac bracket by $-i [.,.]$ leads to the well-known Virasoro
algebra with zero central charge \cite{Green:1987sp}
\eq{[L_k,L_m] = (k-m)L_{k+m}\,.}{eq:Virasoro}

\section{BRST Gauge Fixing}\label{sec:BRST}

The path integral for a system with gauge symmetries is in general
ill-defined. In order to obtain a well-defined path integral one has
to explicitly break gauge invariance by means of constructing an
appropriate gauge fixed action and afterwards restore gauge
independence of the correlation functions of physical fields by
enlarging the phase space with unphysical ghost and antighost fields.

In order to obtain a gauge fixed action, we use the general method of
Batalin, Vilkovisky and Fradkin
\cite{Fradkin:1975cq,Batalin:1977pb,Fradkin:1978xi} based on the BRST
symmetry \cite{Henneaux:1992,Weinberg:1995II}. We first construct the
BRST charge $\Om$. The system has three gauge symmetries generated by
the $G_i$, so the phase space has to be enlarged by three pairs of
ghosts and antighosts
\eq{(c_i, p_i^c)\quad i=1,2,3\,,}{eq:ghosts}
and equipped with a graded Poisson bracket obeying the
(anti)commutation relations \eqref{poisson} and
\eq{\spoiss{c^i}{{p_j^c}'} = - \de^i_j \de(x-x')}{eq:gPBghost}
for the (anti)ghosts. The Dirac bracket is still defined as in
\eqref{diracbracket}, but now with the Poisson bracket of the extended
phase space. One important point is that the ghosts themselfes do not
give rise to new second class constraints, thus the matrices
\eqref{Calphabeta}, \eqref{eq:Calbeinv} are the same as for the gauge
invariant system.  The BRST charge acts on functions on the enlarged
phase space through the Dirac bracket,
\eq{(\Om  F)(q,p,Q,P,c,p^c) := \dirac{\Om}{F}\,,}{eq:BRSTaction}
and has to be nilpotent,
\eq{\Om^2 F = 0\,.}{eq:BRSTnilp}
By virtue of the graded Jacobi identity \eqref{eq:gPBJacobi} this is
equivalent to
\eq{\dirac{\Om}{\Om} = 0\,.}{eq:brstnilpot}
Furthermore $\Om$ is required to act on functions depending only on the
original phase space variables $(q^i,p_i,Q^\al,P_\al)$ like the
original gauge transformations, i.e.  through the generators $G_i$,
and to have ghost number one. This means that one assigns $\gh
(c^i)=+1$ and $\gh (p_j^c) = -1$ for ghosts and antighosts,
respectively, and zero to all other phase space variables. The ghost
number is additive for products of field monomials, $\gh(AB) = \gh(A)
+ \gh(B)$.  This leads to the Ansatz
\eq{\Om = \underbrace{c^i G_i}\limits_{\Om^{(0)}} + {\rm higher\; ghost\; terms}\,.}{eq:BRSTAnsatz}
The first term in the above Ansatz gives (cf. \eqref{eq:structfunc})
\eq{\dirac{\Om^{(0)}}{{\Om^{(0)}}'} = c^i c^j {C_{ij}}^k G_k \de(x-x')\,.}{eq:BRST00}
The next higher term with total ghost number one is 
$$\Om^{(1)} = \frac{1}{2}c^i c^j {C_{ij}}^k p_k^c\,,$$
which cancels the contribution from the zeroth order,
\eq{\dirac{\Om^{(0)}}{\Om^{(1)'}} + \dirac{\Om^{(1)}}{\Om^{(0)'}} = - \dirac{\Om^{(0)}}{\Om^{(0)'}} + D^{(1)}\,.}{eq:BRSTcancel}
Detailed calculations can be found in App. \ref{app:B3}. The
homological perturbation series terminates at Yang-Mills
level\footnote{In Yang-Mills theory the structure functions ${f_{ab}}^c$ are just
  constants and thus the construction must necessarily terminate at
  this order, because $\spoiss{{f_{ab}}^c}{{f_{de}}^f} = 0$.}
\eq{\dirac{\Om^{(1)}}{\Om^{(1)'}} = - D^{(1)}\,,}{eq:BRSTterminate}
not because the structure functions commute with themselfes but rather
because the term containing the bracket of ${C_{ij}}^k$ with itself is
quartic in the ghosts $c^i$, while the system contains only three
anticommuting ghosts \eqref{eq:ghosts}. The full BRST charge thus
reads
\beq
\label{eq:BRSTcharge} \Om = c^i G_i + \frac{1}{2}c^i c^j {C_{ij}}^k p_k^c\,.
\eeq
Constructed in this way it is unique up to canonical transformations
of the extended phase space \cite{Henneaux:1992}.

One now uses the theorem that BRST invariant functionals with total
ghost number zero are sums of a BRST closed and a BRST exact part
\cite{Weinberg:1995II}. The gauge fixed Hamiltonian density should
thus be of form
\beqas
{\cal H}_{gf} = {\cal H}_{BRST} + \dirac{\Om}{\Psi}\,.
\eeqas
Choosing the gauge fixing fermion \cite{Grumiller:2001ea}
\eq{\Psi = p_2^c}{eq:gaugefermion}
and ${\cal H}_{BRST} = 0$ leads to Eddington-Finkelstein gauge
\eq{(\om_0,e_0^-,e_0^+)
  = (0,1,0)\,,}{eq:EFgauge}
and to a gauge fixed Hamiltonian density (compare with \eqref{constrainedhamiltonian})
\eq{{\cal H}_{gf} = \dirac{\Om}{\Psi} = -G_2 - {C_{2i}}^k c^i p_k^c\,.}{eq:gfhamiltonian}
The gauge fixed Lagrangian is obtained through Legendre transform in the extended phase space,
\begin{align}\nonumber
{\cal L}_{gf} & =  \dot{Q}^\al P_\al + \dot{q}^i p_i + p_i^c \dot{c}^i - {\cal H}_{gf} \\
              & =  \dot{Q}^\al P_\al + \dot{q}^i p_i + G_2 + p_k^c {M^k}_l c^l\,, \label{eq:gflagrangian}
\end{align}
and contains the Faddeev-Popov operator
\beq\label{eq:FPoperator}
M = \left(
\begin{array}{lll}
\partial_0 & 0 & \pd{\cV}{X} - \left(g h' - \frac{h}{f} f'g'\cdot (\chib\chi)\right) \\
-1 & \partial_0 & \pd{\cV}{X^+} \\
0 & 0 & \partial_0 + \pd{\cV}{X^-}
\end{array}\right)\,.
\eeq
The gauge choice eq.~\eqref{eq:EFgauge} still admits for some residual
local Lorentz transformations and diffeomorphisms with parameters of
the form
\eq{\ga=\ga(x^1)\,,\quad \xi^0 = x^0 \ga(x^1) + f(x^1)\,,\quad \xi^1 = \xi^1(x^1)\,,}{eq:residualsmall}
as can be seen from the infinitesimal transformations of the gauged
fields
\beqa
0 & = & \de \om_0 = - \partial_0 \ga + \om_\nu \partial_0 \xi^\nu\\
0 & = & \de e_0^- = - \ga + e_\nu^- \partial_0 \xi^\nu \\
0 & = & \de e_0^+ = e_\nu^+ \partial_0 \xi^\nu\,.
\eeqa
Non-infinitesimally these transformations are
\eq{\ga =\ga(x^1)\,,\quad {x^0}' = e^{-\ga(x^1)}(x^0 - g(x^1))\,,\quad
  {x^1}' = {x^1}'(x^1)\,.}{eq:residuallarge}


\chapter{Nonperturbative Quantization of Geometry}\label{ch:nonpert}

In this chapter the path integral over the ghost sector $(c^i,p_j^c)$ and the
geometric variables $(q^i,p_j)$ is evaluated nonperturbatively. The result is an
effective action still depending on the fermions, which are quantized
with perturbative methods in the next chapter.

\section{Ghosts and Second Class Constraints}

First, for later evaluation of correlation functions containing the
geometric variables $(q^i,p_j)$, one couples them and the fermion
field to external sources
\eq{\cL_{\rm src} = J^ip_i + j_i q^i + \etab\chi + \chib\eta\,.}{eq:sourcelagrangian}
The generating functional for correlation functions is formally given by the path integral
\eq{Z[J,j,\eta] = \cN \int \cD\mu[Q,P,q,p,c,p^c] e^{i \int \extd^2x (\cL_{gf} + \cL_{\rm src}) }}{eq:genfunct}
with the actions \eqref{eq:gflagrangian} and \eqref{eq:sourcelagrangian},
and the measure
\eq{\cD\mu[Q,P,q,p,c,p^c] = \cD\mu'[Q]\; \prod\limits_{i=1}^3 \cD p_i
  \cD q^i \prod\limits_{\al=0}^3 \cD P_\al \delta(\Phi_\al)
  \prod\limits_{j=1}^3 \cD c^j \cD p^c_j\,.}{eq:PImeasure}
$\cN$ is a normalization factor. The delta functional in the measure
\cite{Henneaux:1992} restricts the integration to the surface defined
by the second class constraints eqs.~\eqref{Phi0}-\eqref{Phi3}, and
the fermion measure $\cD\mu'[Q]$ will be specified later.

The ghost integration can be carried out immediately, yielding the
functional determinant of the Faddeev-Popov operator
\eqref{eq:FPoperator}
\eq{\Delta_{\Phi\Pi} = \mathrm{Det}(\partial_0^2(\partial_0 + U(X)X^+))\,.}{eq:FPdet}
Integration of the fermion momenta $P_\al$ is trivial because of the
delta functionals in \eqref{eq:PImeasure} and the $P_\al$-linearity of
the second class constraints \eqref{Phi0}-\eqref{Phi3}, yielding an
effective Lagrangian
\eq{\cL_{\rm eff}^{1} = p_i \dot{q}^i + G_2 + \frac{i}{\sqrt{2}}f(p_1)\left[q^3(\chid_0 \lrpd{0} \chi_0) - q^2(\chid_1 \lrpd{0} \chi_1) \right] + \cL_{\rm src}\,.}{eq:Leff1}

\section{On the Path Integral Measure}

Before integrating over $(q^i,p_i)$, one has to specify the
integration measure for the matter fields.  The reason is that for
retaining general covariance of the measure
\cite{Fujikawa:1987ie,Toms:1986sh,Basler:1991st} and thus preventing
the diffeomorphism invariance of the classical field theory from
acquiring an anomaly in the quantum theory, the matter fields in the
integration measure have to be multiplied with appropriate powers of
$\sqrt{-g}$, which in Eddington-Finkelstein gauge \eqref{eq:EFgauge}
is just $q^3 = e_1^+$. For the gauge-fixed path integral this means
that the measure has to be invariant under those BRST transformations
that generate general coordinate transformations
\cite{Fujikawa:1987ie}. As was proposed in \cite{Kummer:1997jj}, we use
this choice of the measure because preserving general covariance also
in the quantum theory is preferable from the physical point
of view and also for subsequent application of known results, e.g. the
derivation of one-loop effects in Ch.~\ref{ch:matter}. The measure can
be formally derived by analogy from the case of a finite dimensional
vector space over complex Grassmann numbers
\cite{Toms:1986sh,Basler:1991st}.  For Dirac fermions in two
dimensions it reads (cf.  eq.~(7.14) in \cite{Basler:1991st} with
diffeomorphism weight $w=0$)
\eq{\cD\mu'[Q] = \prod\limits_x [-g]^{-1} \cD\chib \cD\chi= \prod\limits_x  [e_1^+]^{-2}\, \cD\chib \cD\chi\,. }{eq:mattermeasure}

This procedure of changing the measure by hand to fit
\eqref{eq:mattermeasure} can be implemented already at the level of
the phase space path integral \eqref{eq:genfunct} where, knowing that
the measure will be fixed by hand anyway, clandestinely a factor $\sqrt{\sdet C_{\al\be}}$ has been
dropped.\footnote{Strictly speaking this should be a functional
  superdeterminant, but the functional part $\Det\, \de(x-y) = 1$ has
  already been separated.} This factor arises (see §16.1.1 of
\cite{Henneaux:1992},~\cite{Henneaux:1994jf}) from rewriting the path
integral over the surface of second class constraints 
as a path integral over the whole phase space, with the aforementioned
delta functionals (cf. eq.~\eqref{eq:PImeasure}) again restricting the
allowed paths to lie in the surface defined by second class
constraints.  The superdeterminant of a supermatrix
$$ M = \left(
  \begin{array}{cc}
    A & B\\
    C & D
  \end{array}
\right)\,,
$$
i.e. a matrix with even entries in the Bose-Bose part A and
Fermi-Fermi part D and odd ones in the Bose-Fermi (Fermi-Bose) parts B
(D), is defined via its supertrace \cite{Henneaux:1992}
\eq{\sdet M = \exp \str \ln M\,,\quad \str M = \tr A - \tr D\,.}{eq:defsdet}
Keeping in mind that $C_{\al\be}$ is a pure Fermi-Fermi matrix and
thus
$$\sdet C_{\al\be} = \exp\big[ - \tr \ln C_{\al\be}\big]= (\det C_{\al\be})^{-1}\,,$$
the measure factor becomes
\eq{\sqrt{\sdet C_{\al\be}} = (4f^4(X)(e_1^+ e_1^-)^2)^{-\frac{1}{2}}\,.}{eq:2ndclassmeasure}
For minimal coupling $f(X) = 1$ and with the gauge choice
\eqref{eq:EFgauge} one arrives at the measure
\eqref{eq:mattermeasure} in configuration space by starting with a
phase space measure
\eq{\prod\limits_x [-g^{00}(x)] \sqrt{\sdet C_{\al\be}(x)} \prod\limits_{\al=0}^3 \cD P_\al}{eq:covfermmomentmeasure}
instead of the one used in eq.~\eqref{eq:PImeasure}. This is also what
one would naively expect by analogy from the case of a real boson
field \cite{Toms:1986sh} on a curved background, where a phase space
measure $\cD\mu[\pi,\phi] = \prod\limits_x
[-g^{00}(x)]^{-\frac{1}{2}}\, \extd\pi(x) \,\extd\phi(x)$ leads to the
correct configuration space measure $\prod\limits_x \sqrt[4]{-g(x)}\,
\extd\phi(x)$ after integration over the canonical momenta. The
measure factor for fermions in the Lagrangian path integral can be
expected to be the inverse of the factors in the bosonic case, i.e.
$[-g(x)]^{-1/4}$ for each real degree of freedom, because the volume
element for Grassmann variables transforms inversely to that one
for normal variables under changes of the basis.  Indeed, the Dirac
spinor has two complex and thus four real degrees of freedom, giving
just the right factor \eqref{eq:mattermeasure}. But the kinetic term
for fermions \eqref{eq:fermkin} is already of first order in the time
derivatives and thus fermions and their conjugate momenta are not
independent in the Hamiltonian formalism. If one treats them as
independent variables they only count as one-half degree of freedom
each, yielding the right factor in
\eqref{eq:covfermmomentmeasure}.

For non-minimal coupling the
question of which measure is the ``right'' covariant one is subtle and
still not completely settled (for a review cf.~e.g.
\cite{Kummer:1999zy}).

The additional measure factor \eqref{eq:mattermeasure} introduces a
nonlinearity that prevents the immediate $q^i$-integration, but it can, following
\cite{Kummer:1998zs}, formally be replaced by integration over an
auxiliary field $F$ and functional differentiating with respect to the
current generating the field factor,
\begin{align}
\label{eq:treatmeasure}
Z[J,j,\eta,\etab] & = \cN \int \cD F \de\left(F - \frac{1}{i} \frac{\de}{\de j_3}\right) \tilde{Z}[F,J,j,\eta,\etab] \\ \label{eq:ztilde}
\tilde{Z}[F,J,j,\eta,\etab] & = \int \cD \chib \cD\chi \prod\limits_x F^{-2}\, \prod\limits_{i=1}^3 \cD p_i \prod\limits_{i=1}^3 \cD q^i \Delta_{\Phi\Pi}\, e^{i \int \extd^2 x \cL_{\rm eff}^{(1)}}\,.
\end{align}

\section{The Remaining Gauge Fields}

After eliminating the nonlinearity in the path integral measure, the $q^i$-linearity of \eqref{eq:Leff1}  (cf. \eqref{G2} and
\eqref{G2g}) allows functional integration\footnote{A partial integration
  in the contact term in \eqref{eq:Leff1} produces a boundary
  contribution $\int\extd^2 x\,
  \partial_0(p_iq^i)$, stemming from the unusual order of integration
  $q\rightarrow p$.
Another one comes from
  \eqref{G2g}, $\int \extd^2 x\,
  \partial_1 X^+$, and is cancelled by the boundary term in
  eq.~\eqref{eq:Giboundary}.}
over the $q^i$, yielding three delta functionals containing partial differential
equations for the $p_i$,
\begin{align}\label{eq:p1}
\partial_0 p_1 & =  j_1 + p_2  \\\label{eq:p2}
\partial_0 p_2 & =  j_2 - \frac{i}{\sqrt{2}} f(p_1) (\chid_1 \lrpd{0} \chi_1) \\\label{eq:p3}
(\partial_0 + U(p_1)p_2)p_3 & =  j_3 + \frac{i}{\sqrt{2}} f(p_1) (\chid_0 \lrpd{0} \chi_0) + h(p_1) g(\chib\chi) - V(p_1)\,.\end{align}
These equations are the equations of motion for the $p^i$ that result 
from the gauge fixed Lagrangian \eqref{eq:gflagrangian} after
supplementing it by sources \eqref{eq:sourcelagrangian} and using the
explicit expressions for the fermion momenta $P_\al$ which follow from
the second class constraints~\eqref{Phi0}-\eqref{Phi3}.

Performing the $p_i$-integration is now equivalent to solving these
equations for given currents $j_i$ and matter fields, and
substituting the solutions $p_i = \hat{B}_i$ back into the effective
action obtained after the last step,
\eq{\cL_{\rm eff}^{2} = J^i \hat{B}_i + \etab\chi + \chib \eta + \frac{i}{\sqrt{2}} f(\hat{B}_1) (\chid_1 \lrpd{1} \chi_1)\,.}{eq:Leff2}
During this integration, the Faddeev-Popov determinant
\eqref{eq:FPdet} also formally cancels, because the differential
operators on the left hand side of
eqs.~\eqref{eq:p1}~-~\eqref{eq:p3} produce a factor
$\Det(\partial_0^2(\partial_0 + U(X)X^+))^{-1}$.

Eqs. \eqref{eq:p1} and \eqref{eq:p2} decouple for minimal coupling
$f(X) = -\kappa = \mathrm{const}.$ such that the solutions are formally
\begin{align}\label{eq:p1min}
p_1 = \Bh_1 & =  \nabla_0^{-1}(j_1 + B_2) + \tilde{p}_1 \\\label{eq:p2min}
p_2 = \Bh_2 & =  \nabla_0^{-1}\left( j_2 + \kappa \frac{i}{\sqrt{2}} (\chid_1 \lrpd{0} \chi_1) \right) + \tilde{p}_2 \\ \nonumber
p_3 = \Bh_3 & =  e^{-\Qh}\left[ \nabla_0^{-1} e^{\Qh} \left( j_3 - V(B_1) + h(B_1)g(\chib\chi) - \kappa \frac{i}{\sqrt{2}} (\chid_0 \lrpd{0} \chi_0) \right)\right. \\ \label{eq:p3min}
          &    \hspace{15mm}+ \left.\tilde{p}_3 \right]\,.
\end{align}
Here
\eq{\Qh(\Bh_1,\Bh_2) := \nabla_0^{-1}(U(\Bh_1)\Bh_2)}{eq:Qreg}
is the regularized analogue of \eqref{eq:wI}. For vanishing source $j_1=0$ it can be expressed (using \eqref{eq:p1}) as
\eq{\Qh(\Bh_1) = \int\limits^{\Bh_1} \extd y U(y)\,,}{eq:Qreg2}
where the unspecified lower integration limit means that
\eqref{eq:Qreg} and \eqref{eq:Qreg2} can differ by a homogeneous
solution of the regularized time derivative $\nabla_0$.  The
quantities $\tilde{p}_i$ are also homogeneous solutions of $\nabla_0$.
Here the prescription of App.~B of \cite{Kummer:1998zs} is used, which
provides an infrared regularization and proper asymptotic behaviour of
the Green functions $\nabla_0^{-1}(x,y)$. The details of the
regularization are, however, not important for what follows. In most
calculations which explicitly use the regularized time derivative like
e.g. \cite{Grumiller:2003mc} the important requirement is that
$\nabla_0^{-1} f(x^1)$ should yield $f(x^1) x^0 + g(x^1)$ after
undoing the regularization, i.e. the integral operator
$\partial_0^{-1}$ should act as an anti-derivate.

For general non-minimal coupling $f(X)$
eqs.~\eqref{eq:p1min}~\&~\eqref{eq:p2min} do not decouple, but still
can be solved order by order with the Ansatz
\eq{p_i = \hat{B}_i = \sum\limits_{n=0}^\infty p_i^{(n)}\ \ i=1,2\,,}{eq:piansatz}
assuming the matter contributions in eq.~\eqref{eq:p2} to be small,
i.e. $p_1^{(n)}$ being of order $n$ in
fermion bilinears. Because the coupling functions $f(X),h(X)$ are of order
of magnitude of the gravitational constant this approximation is valid
for systems with total energies several orders of magnitude smaller
than the Planck scale. Quantum gravity effects in few-particle
scattering at energies accessible in present day particle accelerators
are thus safely described in this approach, but the approximation
breaks down either for scattering at the Planck scale or for
macroscopic matter accumulations like macroscopic black holes. In
these cases a nonperturbative solution of eqs.~\eqref{eq:p1} and
\eqref{eq:p2} will be needed, which is only available for minimal
coupling (eqs.~\eqref{eq:p1min} and \eqref{eq:p2min}).

The lowest order contributions $p_i^{(0)}$ are then given by
eqs.~\eqref{eq:p1min}~\&~\eqref{eq:p2min} with $\ka = 0$. For linear
dilaton coupling $f(X) = X$, the higher order terms are obtained through the
recursion relations
\begin{align}\label{eq:p2linear}
p_2^{(n)} & =  -\frac{i}{\sqrt{2}}\nabla_0^{-1}\left( p_1^{(n-1)}(\chid_1 \lrpd{0} \chi_1)\right) \quad n \ge 1 \\\label{eq:p1linear}
p_1^{(n)} & =  \nabla_0^{-1} p_2^{(n)} \,.
\end{align}
For general non-minimal couplings one needs to know the functional
form of $f(X)$ to use this approach. A Taylor expansion around the
zeroth term of eq.~\eqref{eq:piansatz} yields
\eq{f(p_1) = \sum\limits_{n=0}^\infty \frac{f^{(n)}(p_1^{(0)})}{n!} \left( \sum\limits_{m=1}^\infty p_1^{(m)}\right)^n.}{eq:fXTaylor}
One now has to truncate the expansion eq.~\eqref{eq:piansatz} at some order $k\ge
1$ to be able to apply the binomial theorem. The zeroth order
contributions are as above and the first order is
\beq\label{eq:general1storder}
p_2^{(1)} = - \frac{i}{\sqrt{2}}\nabla_0^{-1}\left( f(p_1^{(0)}) (\chid_1 \lrpd{0} \chi_1)\right)\,,\quad p_1^{(1)} = \nabla_0^{-1}p_2^{(1)} \,.
\eeq
Higher orders can be read off order by order from the equations
\beqa\label{eq:p1general}
 \sum\limits_{m=2}^{k+1} \nabla_0 p_1^{(m)} & = & \sum\limits_{m=2}^{k+1} p_2^{(m)} \\\nonumber
\sum\limits_{m=2}^{k+1} \nabla_0 p_2^{(m)} & = &  - \frac{i}{\sqrt{2}} \sum\limits_{n=1}^\infty  \frac{f^{(n)}(p_1^{(0)})}{n!} \left( \sum\limits_{m=1}^k p_1^{(m)}\right)^n (\chid_1 \lrpd{0} \chi_1) \\\nonumber
& = & - \frac{i}{\sqrt{2}} \sum\limits_{n=1}^\infty \frac{f^{(n)}(p_1^{(0)})}{n!} \sum\limits_{l_1=0}^n \sum\limits_{l_2=0}^{l_1}\dots\sum\limits_{l_{k-2}=0}^{l_{k-1}} {n \choose l_1} {l_1 \choose l_2}\dots {l_{k-2}  \choose l_{k-1}} \cdot \\\label{eq:p2general}
 && \hspace{-2cm}\underbrace{\cdot \left( p_1^{(1)} \right)^{n-l_1} \left( p_1^{(2)} \right)^{l_1-l_2}\dots\left( p_1^{(k-1)} \right)^{l_{k-2}-l_{k-1}} \left( p_1^{(k)} \right)^{l_{k-1}} \cdot (\chid_1 \lrpd{0} \chi_1)}\limits_{{\rm of \  order\ }n + l_1 + l_2 + \dots + l_{k-1} + 1} \,.
\eeqa
For a given order $m$ this yields restrictions
\beq\label{eq:powerrestrict}
m = n + l_1 + l_2 + \dots + l_{k-1} + 1,\quad n \ge l_1 \ge l_2 \ge \dots \ge l_{k-1} \ge 0
\eeq
which can be used to pick the terms of order $m$ on the right hand
side of eq.~\eqref{eq:p2general}. For a given $m$ the highest order in
$p_1$ that can contribute to the equation determining $p_2^{(m)}$ is
contained in the term proportional to
$$ f^{(1)}(p_1^{(0)}) p_1^{(m-1)} (\chid_1 \lrpd{0} \chi_1) \,,$$
corresponding to $n=l_1=\dots=l_{m-2} = 1, l_{m-3} = \dots = l_{k-1} =
0$. Thus the $m$th order $p_{1/2}^{(m)}$ only depend on
the previous orders, and this expansion allows to solve
eqs.~\eqref{eq:p1}~and~\eqref{eq:p2} order by order in the fermion
bilinears. The $(k+1)$th order can also be calculated from
eqs.~\eqref{eq:p1general}~and~\eqref{eq:p2general}, and at the end the
truncation of eq.~\eqref{eq:piansatz} turns out to be a technical
assumption needed for applicability of the binomial theorem and, of
course, for avoiding convergence issues of this
expansion. Nonetheless the right hand side of eq.~\eqref{eq:p2general}
contains all the terms necessary to calculate $p_{1/2}$ to arbitrary
high orders.

Eq.~\eqref{eq:p3} can always be solved, yielding
(for $j_1 = 0$ and with eq.~\eqref{eq:Qreg2})
\begin{multline}
p_3 = \hat{B}_3 = e^{-Q(\hat{B}_1)}\Big[ \nabla_0^{-1} e^{Q(\hat{B}_1)} \Big( j_3 - V(\hat{B}_1) + h(\hat{B}_1)g(\chib\chi) \\
+ \frac{i}{\sqrt{2}} f(\hat{B}_1) (\chid_0 \lrpd{0} \chi_0) \Big) + \tilde{p}_3 \Big]\,. \label{eq:p3general}
\end{multline}

Although the treatment of general dilaton couplings is possible, the
physically most interesting cases of Einstein-Hilbert gravity with in
four dimensions minimally coupled scalars, fermions and Yang-Mills
fields yield upon spherical reduction kinetic terms with minimal or
linear dilaton coupling \cite{Balasin:2004gf}, so that these two cases
seem to be the most important ones. This is easily understood: Because
the spherical symmetric Ansatz
\eq{(\extd s)^2 = h_{\al\be}\extd x^\al \extd x^\be - X(x^\al) (\sin^2 \theta \extd \phi^2 + \extd \theta^2)}{eq:sphericalmetric}
reduces the four-dimensional volume element $\extd^4 x \sqrt{-g}$ upon
integration over the angular part to $\propto \extd^2 x X(x^\al)
\sqrt{-h}$, spherical reduction of in four dimensions minimally coupled
matter will in general lead to a linear dilaton coupling in two
dimensions.  One important exception could be 2D Type 0A/0B string
theory (cf. the penultimate model in table~\ref{tab:1}), if it could
be generalized to include the same nonperturbative corrections that
are already included in the exact string black hole (ESBH)
\cite{Grumiller:2005sq,Grumiller:2006rc}. For the ESBH the first order
gravity dilaton field $X$ and the string dilaton $e^{-2\phi} = \ga B$,
with $B$ being an auxiliary field, are related by
\eq{X = \ga + {\rm arcsinh} \ga\,.}{eq:0A0Bdil}
The question now stands whether the string tachyon nonperturbatively
couples to the first order gravity dilaton $X$, as it does in the
perturbative limit (cf. §3.1 of \cite{Grumiller:2006rc}), or to the
string dilaton $\ga$. In the first case the coupling is linear, but
in the second case it gets nonlinear corrections. Also some "two
dilaton theories" \cite{Grumiller:2000wt} show nonlinear coupling to
matter fields.

\section{Ambiguous Terms}\label{sec:ambiguous}

Eq.~\eqref{eq:Leff2} as it stands is not the whole effective action,
but contains an ambiguity \cite{Kummer:1998zs,Grumiller:2000ah}
arising from the source terms $J^i \hat{B}_i$. In expressions like
$\int J \nabla^{-1} A$ the regularized inverse derivative, which
stands for an integral operator, acts after a partial integration and changing the order of integration on the source $J$ and in this way giving rise to another homogeneous contribution
$\int \tilde{g} A$, whereas the homogeneous functions in $A$ have
already been made explicit in the solutions $\hat{B}_i$. The action
thus has to be supplemented by three terms
\begin{align}\nonumber
  \cL_{amb} & = \sum\limits_{i=1}^2 \tilde{g}_i
  K_i(\nabla_0^{-1},(\chid_1\lrpd{0}\chi_1),j_1,j_2)
  \\\label{eq:ambiguous} & \hspace{5mm}+ \tilde{g}_3 e^{\hat{Q}} \left(
    j_3 - V(\hat{B}_1) + h(\hat{B}_1)g(\chib\chi) + f(\hat{B}_1)
    \frac{i}{\sqrt{2}} (\chid_0 \lrpd{0} \chi_0) \right)\,.
\end{align}
The quantities $K_i$ can be read off from the solutions
$\hat{B}_{1/2}$ up to the desired order in matter contributions.  The
first two terms in general do not generate any new couplings of the
matter fields besides the ones generated by the $-\tilde{g}_3
w'(\hat{B}_1) = -\tilde{g}_3 e^{\hat{Q}(\hat{B}_1,\hat{B}_2)}
V(\hat{B}_1)$ and thus can be omitted unless $w'(X) = {\rm const.}$ In
that case they become important and, for example, yield corrections to
the specific heat of the CGHS black hole (cf. the third model in
table~\ref{tab:1}) when coupled to a scalar field
\cite{Grumiller:2003mc}, changing it from a meaningless infinity to a
positive finite value.

That these ambiguous terms are necessary and cannot be omitted is also
obvious from \eqref{eq:Leff2}.  We know that classically the fermions
couple to the metric determinant $\sqrt{-g}=q^3$ and this should still
be the case after integration of the geometric degrees of freedom,
where the coupling is to an effective background defined by the $j_i$
and the matter fields.  After setting $J^i=0$ the action
\eqref{eq:Leff2} is independent of $j_3$ and thus because of
\eqref{eq:treatmeasure} containing a factor $F^{-2}$ and the delta
functional therein setting $F=0$ the whole partition function would be
ill-defined without the ambiguous terms.

Furthermore, first order gravity in two dimensions without matter is
locally quantum trivial \cite{Kummer:1996hy}, i.e. the effective
action expressed in terms of the mean fields $\langle q^i \rangle =
\de S_{\rm eff}/\de j_i$ and $\langle p_i \rangle = \de S_{\rm
  eff}/\de J^i$ is, up to a boundary term, the classical action
\eqref{eq:FOG} in Eddington-Finkelstein gauge. Thus no local quantum
corrections appear in first order gravity and all eventual quantum
effects are encoded in the boundary part and are of global nature. The
mean fields then fulfill the classical equations of motion. In the
matterless case, i.e.  dropping the $\cD\chib\cD\chi$-integration and
the covariant measure factor and setting $f(X)=h(X)=0$, the correlator
\beq
\langle e_1^+ \rangle = \left.\frac{1}{iZ[J=j=0]}\frac{\de}{\de j_3} e^{i\int\extd^2 x(\cL_{\rm eff}^{2} + \cL_{amb})}\right|_{j_i=J^i=0} = \tilde{g}_3 e^{Q(B_1[j_i=J^i=0])}
\eeq
is indeed just the classical solution for $e_1^+$, cf.
eq.~\eqref{eq:sole+}, in the gauge \eqref{eq:EFgauge} (with $\extd x^1
= X^+ \extd f$ and $\tilde{g}_3 = 1$). Thus for obtaining the right
classical solutions in the theory without matter the third term in
\eqref{eq:ambiguous} is indispensable. This example also shows that
the homogeneous functions $\tilde{g}_i$ are fixed by imposing
asymptotic conditions on the expectation values of the geometric
fields $q^i$.

The occurrence of the ambiguity is also connected to the unusual order
of integrating first over the canonical coordinates $q^i$ and then
over the momenta $p_i$. As shown in \cite{Haider:1994cw}, for the
special case of the Katanaev-Volovich model (the fourteenth entry in
table~\ref{tab:1}) without matter the path integration can be carried
out in "normal" order, i.e. first over $p_i$ and then over $q^i$,
without introducing sources $J^i$, yielding an effective action
exactly of the type of the last term of eq.~\eqref{eq:ambiguous}.

After quantizing the whole ghost and geometry sector, the generating
functional reads
\beqa
\label{eq:z2}
Z[J,j,\eta,\etab] & = & \cN \int \cD F \de\left(F - \frac{1}{i} \frac{\de}{\de j_3}\right) \tilde{Z}[F,J,j,\eta,\etab] \\\label{eq:ztilde2}
\tilde{Z}[F,J,j,\eta,\etab] & = & \int \cD \chitb \cD\chit \left. e^{i \int \extd^2 x (\cL_{\rm eff}^{2} + \cL_{amb})}\right|_{\chi = F^{-\frac{1}{2}}\chit}\,,
\eeqa
with $\cL_{\rm eff}^{(2)}$ from \eqref{eq:Leff2} and $\cL_{\rm amb}$ from
\eqref{eq:ambiguous}.  It should be emphasised that $Z$ includes all
gravitational backreactions, because the auxiliary field $F$ upon
integration is equivalent to the quantum version of $e_1^+$.

\section{Conformal Properties of the Effective Action}\label{sec:confEF}

The action of conformal transformation $g_{\mu\nu}=\Om^2 \tilde
{g}_{\mu\nu}$ (or equivalently $g_{\mu\nu} \mapsto \Om^2 g_{\mu\nu}$)
after fixing the gauge and specifying asymptotic values of the $p_i$
is slightly non-trivial and requires some discussion. This is a
necessary pre-requisite to understand the conformal properties of the
vertices and the scattering matrix. By assumption it will be required
that the asymptotic values $\tilde{p}_i$ and the gauge fixing
conditions \eqref{eq:EFgauge} are invariant. This implies in
particular that neither $e_0^-$ nor $X^+$ transform and that $e_1^+$
has the same conformal weight as the metric. Furthermore, the
geometric part of the conserved quantity
eq.~\eqref{eq:conservedquantity} is required to be conformally
invariant. This leads to different conformal weights (listed in
table~\ref{tab:conf}) as compared to the situation before gauge
fixing, eq.~\eqref{eq:FOGconftrans}. The conformal weight $w(A)$ of a
field monomial $A$ is defined (if it transforms homogeneously) by $A =
\Om^{w(A)} \tilde{A}$. Conformal weights thus add for products of
field monomials, $w(AB) = w(A) + w(B)$.
\begin{table}[h]
\centering
\begin{tabular}{|c|c|p{4cm}|c|c|}\hline
Weight 2 & Weight 1 & Weight 0 & Weight -1 & Weight -2 \\ \hline
$g_{\mu\nu}$, $e_1^+$, $J^3$ & $\eta_1$ & $e_0^\pm$, $e_1^-$, $X$, $X^+$, $\tilde{p}_i$, $\tilde{g}_i$, $\chi_1$, $J^{1/2}$, $j_{1/2}$, $\eta_0$, $A^+$ & $\chi_0$ & $X^-$, $j_3$, $A^-$ \\ \hline
\end{tabular}
\caption{Conformal weights for Eddington-Finkelstein gauge}
\label{tab:conf}
\end{table}

The gauge fixed spin-connection transforms inhomogeneously,
\begin{equation}
  \label{eq:fer1}
  \om_0\to\tilde{\om}_0=\om_0=0\,,\quad \om_1\to\tilde{\om}_1=\om_1+(X^+e_1^- + X^-e_1^+)\frac{\extd\,\ln{\Om}}{\extd X}\,,
\end{equation}
and the dilaton potentials transform as in the gauge invariant theory,
cf. sec.~\ref{ssec:conf}. With this choice the effective action
\eqref{eq:Leff2} and \eqref{eq:ambiguous} is 
conformally invariant at tree-level in the matter fields, i.e. before
perturbatively performing the path integration over $\chi$ and thus
taking into account quantum corrections from the fermions.


\chapter{Perturbative Treatment of the Matter Fields}\label{ch:matter}

After having obtained the effective action \eqref{eq:Leff2} and
\eqref{eq:ambiguous} the remaining matter integration in
\eqref{eq:ztilde2} is carried out perturbatively in this chapter,
following the treatment of the scalar case \cite{Kummer:1998zs}. One
first splits the effective action \eqref{eq:Leff2} and
\eqref{eq:ambiguous} into terms independent of the fermions, in those
quadratic in the spinor components and in higher order terms
summarized in an interaction Lagrangian $\cL_{int}$. The solutions of
eqs.~\eqref{eq:p1}~and~\eqref{eq:p2} up to quadratic fermion terms are
given by the zeroth order solutions $B_i$ (eqs. \eqref{eq:p1min} and
\eqref{eq:p2min} with $\ka=0$) plus the first order contribution
eqs.~\eqref{eq:general1storder}. The expansion \eqref{eq:piansatz}
then reads
\begin{align}\label{eq:ph1}
\Bh_1 & =  B_1 -\frac{i}{\sqrt{2}} \nabla_0^{-2} (f(B_1) (\chid_1 \lrpd{0} \chi_1)) + \cO(\chi^4)\\\label{eq:ph2}
\Bh_2 & =  B_2 -\frac{i}{\sqrt{2}} \nabla_0^{-1} (f(B_1) (\chid_1 \lrpd{0} \chi_1)) + \cO(\chi^4)\,.
\end{align}
Expanding\footnote{The space-time points where the functions are
  evaluated at are denoted in subscript. Double subscripts like
  $G_{xy}$ mean $G(x,y)$, and $\int_x = \int \extd^2 x$.}
eq.~\eqref{eq:Qreg}
\begin{align}
\Qh(\hat{B}_1,\hat{B}_2) & =  Q(B_1,B_2) - \frac{i}{\sqrt{2}} \int_y G_{xy} f(B_{1y})(\chid_1 \lrpd{0} \chi_1)_y \label{eq:Qh} + \cO(\chi^4)\\
G_{xy} & =  \int_z \nabla_{0xz}^{-1}[U'_z B_{2z} \nabla_{0zy}^{-2} + U_z\nabla_{0zy}^{-1}]\, \label{eq:Gxy}
\end{align}
and $\Bh_3$ (eq.~\eqref{eq:p3general}) up to quadratic terms in the
fermions yields
\begin{align}\nonumber
\Bh_3 & =  B_3 + \frac{i}{\sqrt{2}} \int_y H_{xy} f(B_{1y}) (\chid_1 \lrpd{0} \chi_1)_y \\\label{eq:Bh3exp}
      &    \hspace{5mm} + e^{-Q_x} \nabla_{0}^{-1} \left( e^{Q} \left( \frac{i}{\sqrt{2}} f(B_{1}) (\chid_0 \lrpd{0} \chi_0) + h(B_{1}) m \chib\chi\right)\right) \\ \nonumber
H(x,y) & =  e^{-Q_x}\int_z \nabla_{0xz}^{-1}e^{Q_z}\left\{ [G_{xy} - G_{zy}] (j_3 - V)_z \right.\\\label{eq:Hxy}
       &    \hspace{21mm} \left.+ V'_z\nabla_{0zy}^{-2}\right\} + \tilde{p}_{3x}e^{-Q_x}G_{xy}\,.
\end{align}
A similar expansion of the ambiguous terms eq.~\eqref{eq:ambiguous}
with $\tilde{g}_1 = \tilde{g}_2 = 0$ yields
\begin{align}\label{eq:Leffexp}
\cL_{\rm eff} & =  \cL_{\rm eff}^{(0)} + \cL_{\rm eff}^{(2)} + \cL_{int} \\\label{eq:Leff0}
\cL_{\rm eff}^{(0)} & =  J^iB_i + \tilde{g}_3 e^Q(j_3-V(B_1))  \\\nonumber
\cL_{\rm eff}^{(2)} & =  \frac{i}{\sqrt{2}} f(B_1) \left[(\chid_1\lrpd{1}\chi_1) - E_1^- (\chid_1\lrpd{0}\chi_1) + F^{(0)} (\chid_0 \lrpd{0} \chi_0) \right] \\ \label{eq:Leff_2}
                &    \hspace{5mm} + F^{(0)} h(B_1) m \chib\chi + \etab\chi + \chib\eta \\\nonumber
E_1^-(x)        & =  \int_y \Big[ J^1_y\nabla_{0yx}^{-2} + J^2_y \nabla_{0yx}^{-1} - J^3_y H_{yx} \\\label{eq:E1-}
& \hspace{1cm}+ \tilde{g}_{3y} e^{Q_y}(G_{yx}(j_3-V)_y - V'_y\nabla_{0yx}^{-2})\Big]\\\label{eq:E1+}
{E_1^+}^{(0)}(x)        &  = e^{Q_x}\left[ \int_y J^3_y e^{-Q_y} \nabla_{0yx}^{-1} + \tilde{g}_{3x} \right] =: F^{(0)}\,.
\end{align}
One recognizes the kinetic term \eqref{eq:Lkin} of fermions on a
curved background in Eddington-Finkelstein gauge \eqref{eq:EFgauge}
with a background metric
\beq \label{eq:EFmetric}g_{\mu\nu} = F^{(0)} \left(
\begin{array}{cc}
0 & 1 \\
1 & 2 E_1^-
\end{array}\right)_{\mu\nu} \,.
\eeq
This background is still the classical one, solely depending on
sources for the geometric variables $(q^i,p_i)$ and the zeroth order 
solutions $B_i$. If taking into account the first two ambiguous terms
in \eqref{eq:ambiguous}, $\cL_{\rm eff}^{(0)}$ and $E_1^-$ acquire
additional contributions
\beq
\cL_{\rm eff}^{(0)} + \tilde{g}_1\left( j_1 + B_2\right) + \tilde{g}_2 j_2
\eeq
and
\beq
E_1^-(x) + \tilde{g}_2(x) + \int_y \tilde{g}_{1y}\nabla_{0yx}^{-1}\,.
\eeq
After redefining the interaction part of the Lagrangian density such
that the background \eqref{eq:EFmetric} depends on the full $E_1^+(x) = F(x) =
\frac{\de}{\de j_3(x)} \int \extd^2 z \cL_{\rm eff}(z)$ instead of its
matter-independent part $F^{(0)}$, i.e.  taking into account
backreactions onto the metric determinant to all orders in the fermion
fields, the generating functional \eqref{eq:ztilde2} becomes
\begin{multline}\label{eq:ztilde3}
\Zt = \exp \left( i\int \extd^2 x \cL_{\rm eff}^{(0)} + \cL_{int}\left[\frac{1}{i} F^{-\frac{1}{2}} \frac{\de^L}{\de \etatb} \right] \right) \times \\
 \times \int \cD\chitb \cD \chit \exp \left(  i \int \extd^2 x  \frac{i}{2}f(B_1) \epsilon_{ab} \tilde{\epsilon}^{\nu\mu} E_\mu^b (\chitb \ga^a \lrpd{\mu}  \chit) + \etatb \chit + \chitb \etat \right) .
\end{multline}
There is one problem to address in connection with this interpretation
of fermions on an effective background: The metric becomes complex if
one chooses the regularisation of the inverse derivatives as in
\cite{Kummer:1998zs}. To retain reality of the effective metric one
could define it to be the solution of the equations of motion of
\eqref{eq:Leff1}, i.e.  eqs.~\eqref{eq:p1}~-~\eqref{eq:p3}, the
equations for $q^i$ 
and the constraints $G_i = 0$, together with appropriate boundary
conditions and for given external sources.  In what follows the
effective background is assumed to be real. When calculating vertices
in sec.~\ref{sec:vertices} below, instead of expanding the effective
action a method (first introduced in \cite{Kummer:1998zs}) that uses
exactly this idea is applied.  Another solution would be to find a
regularisation which gives a real integral kernel $\nabla_0^{-1}$.

\section{One-Loop Effects and Bosonization}\label{sec:oneloop}

In this section the fermions are assumed to be massless and minimally
coupled. A self interaction term $g(\chib\chi) = \la(\chib\chi)^2$ can
be rewritten by introducing an auxiliary vector potential, 
\eq{S_{SI}[\chi,g,A] = \frac{\la}{2} \int\extd^2 x \left( F A_a^2 + 2 A_a \chitb \ga^a \chit \right)\,,}{eq:rewrThirring}
which is integrated over in the path integral. The last term will be
absorbed into the Dirac operator, resembling minimal coupling of
fermions to the vector potential. After replacing $\partial_\mu$ in
\eqref{eq:ztilde3} with the covariant derivative $\nabla_\mu$
containing the metric compatible and torsion-free spin connection
(which drops out in two dimensions anyway), partially integrating the
kinetic term in \eqref{eq:ztilde3}\footnote{Or in other words using
  the fact that the Dirac operator $\Dir$ in Minkowski space is
  self-adjoint with respect to the Dirac inner product with $\chib =
  \chi^\dagger \ga^0$.}, completing the square and evaluating the
Gaussian integral over $\chi$ and $\chib$ yields
\begin{multline}
\Zt = \exp\left( { i\int \extd^2 x \cL_{\rm eff}^{(0)} + \cL_{int}\left[-i F^{-\frac{1}{2}} \frac{\de^L}{\de \etatb} \right]} \right) \times \\
\times \int \cD A \, \Det \Dir \, \exp\left({i\int \extd^2 x \left( \frac{\la}{2} F A_a^2 - \etatb \Dir^{-1} \etat\right)}\right) \label{eq:ztilde4}
\end{multline}
with the Dirac operator
\eq{\Dir = i E_a^\mu \ga^a (\nabla_\mu -i \la A_\mu)\,,\quad \nabla_\mu = \partial_\mu - \frac{1}{2} \om_\mu \ga_\ast\,.}{eq:diracop}
Conformal transformations are homogeneous transformations of the
metric $g_{\mu\nu}\mapsto \Om^2 g_{\mu\nu}$. The Dirac operator
without vector potential transforms homogeneously with weight $-3/2$
when acting on a spinor which itself has weight $-1/2$, i.e.
$\Dir[A=0]\chi \mapsto \Om^{-\frac{3}{2}}\Dir[A=0]\chi$, and the
kinetic term
$$(\chi,\Dir \chi) = \int \extd^2 x \sqrt{-g}\chib \Dir \chi$$
is conformally invariant. To retain this property in the case of
non-vanishing vector potential, $A_a$ needs to have conformal weight
$-1$ (coming from $E^\mu_a$), while $A_\mu$ has weight 0. With
Eddington-Finkelstein gauge \eqref{eq:EFgauge} fixed the appropriate
weights are $0$ for $A^+$ and $-2$ for $A^-$ and $A^\mu$.

The determinant of the Dirac operator is most easily calculated using
heat kernel methods \cite{Vassilevich:2003xt} in Euclidean space.
There $\Dir$ is essentially self-adjoint \cite{Friedrich:1997} with
respect to the ordinary inner product on spinor space
$$(\chi,\psi) = \int \extd^2 x \sqrt{-g} \chi^\dagger \psi\,.$$
The square of the Dirac operator is then of Laplace type (cf.
eqs.~(3.27) in \cite{Vassilevich:2003xt} with $A_\mu^5=0$),
\eq{D = \Dir^2 = -(g^{\mu\nu} \nabla_\mu \nabla_\nu + E)\,,}{eq:Diracsquared}
with
\eq{E = -\frac{R}{4} + \frac{i\la}{2} \ga_\ast \epsilon^{\mu\nu}F_{\mu\nu}}{eq:DiracsquaredE}
and $F_{\mu\nu} = \partial_\mu A_\nu - \partial_\nu A_\mu$ denotes the
electromagnetic field strength. Conventions applied for Euclidean
space can be found in App.~\ref{conventions}.

\subsection{Zeta Function Regularization and the Heat Kernel}

For deriving one-loop results for classically conformal and chiral
invariant field theories the application of heat kernel methods and
zeta function regularization \cite{Vassilevich:2003xt} is most
convenient. The goal of this subsection is to calculate the one-loop
effective action
\eq{W_{\rm 1loop} = - \ln \Det \Dir = - \frac{1}{2} \ln \Det D = -\frac{1}{2} \Tr(\ln D)\,.}{eq:W1}
For positive eigenvalues $\la$ of $D$ the formal identity
\eq{\ln \la = - \int\limits_0^\infty \frac{\extd t}{t} e^{-t\la}}{eq:lntoexp}
can be used to express the effective action
\eq{W = \frac{1}{2} \int\limits_0^\infty \frac{\extd t}{t} K(t,D)}{eq:W2}
in terms of the heat kernel
\eq{K(t,f,D) = \Tr\left(f e^{-tD}\right)\,,\quad K(t,D) = K(t,1,D)\,.}{eq:heatkernel}
Eq.~\eqref{eq:lntoexp} is only true up to an infinite contribution
that does not depend on $\la$, as can be seen from the Laplace
transform (using eq.~\eqref{eq:Gaexpand})
\beq\label{eq:Tnlaplace}
\int\limits_0^\infty t^{s-1} e^{-t\la}\extd t = \cL[t^{s-1}] = \frac{\Ga[s]}{\la^s} = \frac{1}{s} - \ga_E - \ln \la + \cO(s)\,.
\eeq
Here $\ga_E$ is the Euler-Mascheroni constant. Extending
\eqref{eq:lntoexp} to zero or negative eigenvalues is only formally
possible, because the integral does not converge at the upper limit.
These eigenvalues correspond to infrared divergences that can be
regularized by e.g. adding a mass to the Laplace type operator, $D
\mapsto D + m^2$. The divergence at the lower limit corresponds to the
ultraviolet range and has to be properly regularized by e.g. zeta
function regularization, i.e. by using the Laplace transformed
effective action \eqref{eq:W2} before taking the limit $s\rightarrow
0$ as the regularized effective action
\eq{W_s = \frac{1}{2} \tilde{\mu}^{2s} \int\limits_0^\infty \frac{\extd t}{t^{1-s}} K(t,D),\quad s\ge 0\,.}{eq:W3}
The quantity $\tilde{\mu}$ is a constant with the dimension of a mass
introduced to keep the dimension of the effective action unchanged.
The regularization is removed in the limit $s\rightarrow 0_+$. The key
result used below is that on manifolds of dimension $n$ without
boundaries (or on manifolds with boundaries and local Neumann,
Dirichlet or mixed boundary conditions) the heat kernel
\eqref{eq:heatkernel} admits an asymptotic expansion for small $t$,
\eq{K(t,f,D) \cong \sum\limits_{k\ge 0} t^{(k-n)/2}a_k(f,D)\,,}{eq:HKexpand}
where the heat kernel coefficients $a_k$ are locally computable from
invariants of the manifold, of the bundle part $E$ of the operator
\eqref{eq:Diracsquared} and possibly more invariants if the smearing
function $f$ itself takes values in some internal gauge group. They
are tabulated in e.g.  \cite{Vassilevich:2003xt}.  The zeta function
for a positive operator $D$ is defined as
\eq{\zeta(s,f,D) = \Tr \left(f\,D^{-s}\right)\,,\quad \zeta(s,D) = \zeta(s,1,D)\,,}{eq:zetadef}
and is related to the heat kernel \eqref{eq:heatkernel} by the Mellin
transform
\beqa
\label{eq:zetaheatkernel}
\zeta(s,f,D) & = & \Ga(s)^{-1} \int\limits_0^\infty \extd t\, t^{s-1}\, K(t,f,D)\\\label{heatkernelzeta}
K(t,f,D)     & = & \frac{1}{2\pi i}\oint \extd s\, t^{-s}\, \Ga(s)\, \zeta(s,f,D)\,,
\eeqa
where the integration contour in the second equation encircles all
poles of the integrand. For operators with negative and zero modes one
should replace $D$ in \eqref{eq:zetadef} with the absolute value $|D|$
and restrict the sum in the trace to non-zero eigenvalues. The zeta
function is regular at $s=0$. Comparing \eqref{eq:HKexpand} with
\eqref{heatkernelzeta} relates the heat kernel coefficients to poles
of the zeta function,
\eq{a_k(f,D) = {\rm Res}_{s=(n-k)/2} (\Ga(s)\zeta(s,f,D))\,,}{eq:HKcoeffzeta}
which becomes clear after recalling that the gamma function has a
simple pole at $s=0$,
\eq{\Ga(s) = \frac{1}{s} - \ga_E + \cO(s)\,.}{eq:Gaexpand}
In particular
\eq{a_n(f,D) = \zeta(0,f,D)}{eq:an}
will be used in the following.  The regularized action \eqref{eq:W3}
can now be expressed as
\eq{W_s = \frac{1}{2} \tilde{\mu}^{2s} \Ga(s)\zeta(s,D)\,,}{eq:W4}
and expanded around $s=0$ with the use of \eqref{eq:Gaexpand}
($\mu^2=e^{-\ga_E}\tilde{\mu}^2$), revealing a pole at $s=0$,
\eq{W_s = \frac{1}{2}\left( \frac{1}{s} + \ln \mu^2 \right)\zeta(0,D) + \frac{1}{2}\zeta'(0,D),}{eq:W5}
that has to be removed by renormalization such that the limit $s
\rightarrow 0_+$ can be safely taken. The remaining part is the
renormalized effective action
\eq{W^{\rm ren} = \frac{1}{2}\zeta'(0,D) + \frac{1}{2}\ln(\mu^2) \zeta(0,D)\,.}{eq:Wren}
The variation of the zeta function can be rigorously derived
\cite{Vassilevich:2003xt} by use of the Mellin transform
eqs.~\eqref{eq:zetaheatkernel},~\eqref{heatkernelzeta} and coincides
with the naive variation
\eq{\de\zeta(s,D) = -s \Tr((\de D)D^{-s-1})\,.}{eq:zetavar}

\subsection{Conformal Anomaly}\label{ssec:confanom}

The energy-momentum tensor in classical field theory with action $S$
on a curved Euclidean background is defined as
\eq{T_{\mu\nu} = \frac{2}{\sqrt{g}} \frac{\de S}{\de g^{\mu\nu}}\,.}{eq:EMtensor}
If the theory is conformally invariant, the variation of the action
has to vanish under arbitrary (infinitesimal) conformal
transformations of the metric $\de g^{\mu\nu} = -2 \de\rho \,
g^{\mu\nu}$,
\eq{0 = \de S = \frac{1}{2} \int \extd^2 x \sqrt{g} T_{\mu\nu}\de g^{\mu\nu} = - \int \extd^2 x \sqrt{g} T_\mu^\mu \de\rho\,,}{eq:EMtensorvar}
and thus the energy-momentum tensor of a classically conformally
invariant field theory is necessarily traceless,
\eq{T_\mu^\mu = T_{\mu\nu}g^{\mu\nu} = 0\,.}{eq:EMtensortraceless}

In the quantum theory this no longer needs to be the case. Defining
the Euclidean quantum effective action as $W = - \ln Z$ and the
expectation value of the energy-momentum tensor as
\eq{\langle T_{\mu\nu} \rangle = \frac{2}{\sqrt{g}} \frac{\de W}{\de g^{\mu\nu}}\,,}{eq:QEMtensor}
the variation of the effective action is related to the trace of the
quantum energy momentum tensor $T_\mu^\mu = g^{\mu\nu} \langle
T_{\mu\nu} \rangle$ like in \eqref{eq:EMtensorvar} with $S$ replaced
by $W$. For the classical action to be conformally invariant, the
square of the Dirac operator has to transform conformally covariant,
$D\rightarrow e^{-2\rho} D$. The infinitesimal variation is $\de D =
-2 (\de \rho)D$, and varying the renormalized effective action
\eqref{eq:Wren} yields (cf. \eqref{eq:zetavar})
\beqa\nonumber
\de W^{\rm ren} & = & \frac{1}{2}\left.\frac{\extd}{\extd s}\right|_{s=0} \de\zeta(s,D)  + \frac{1}{2} \ln(\mu^2) \de\zeta(0,D)\\\nonumber
                & = & \zeta(0,\de\rho,D)\\\label{deltaWrenconf}
                & = & a_2(\de\rho,D)\,.
\eeqa
Comparing with eq.~\eqref{eq:EMtensorvar} for the quantum effective
action shows that the trace of the quantum energy-momentum tensor
eq.~\eqref{eq:QEMtensor} is proportional to the second local heat
kernel coefficient whose structure \cite{Vassilevich:2003xt} is
especially simple, yielding with eq.~\eqref{eq:DiracsquaredE} and
$\tr_{{\mathbb C}^2} {\mathbf{1}} = 2$
\eq{T_\mu^\mu = -a_2(x,D) = - \frac{1}{4\pi}
  \tr_{\complexc^2}\left(\frac{R}{6} + E\right) =
  \frac{R}{24\pi}\,.}{eq:confanomaly}
The field strength term in $E$ does not contribute because of the
spinor trace ${\rm tr_{\complexc^2}}(\ga_\ast) = 0$. Notably, the
conformal anomaly for a Dirac fermion in two dimensions has the same
value as for one real boson \cite{Vassilevich:2003xt}, and can
immediately be used to integrate (with initial condition $W[g_{\mu\nu}
= \eta_{\mu\nu}] = 0$) the variation of the 1-loop effective action
(now in Minkowski space) to the nonlocal Polyakov action
\cite{Polyakov:1981rd}
\eq{W_{\rm Pol}[g] = \frac{1}{96\pi}
  \int\limits_{\cM} \extd^2 x \sqrt{-g} R \frac{1}{\Delta}
  R\,,}{eq:Polyakovaction}
with the Laplacian on functions ($\extd^\dagger f(x) = 0$,
cf.~\eqref{eq:ddagger})
\eq{\Delta f = \extd^\dagger \extd = \ast\, \extd \ast \extd = \frac{1}{\sqrt{-g}} \partial_\mu(\sqrt{-g}g^{\mu\nu}\partial_\nu f)\,,}{eq:Laplacian}
and Green's function defined as
\eq{\Delta_x \Delta^{-1}(x,y) = \de(x-y)\,.}{eq:Greensfunction}

\subsection{Chiral Anomaly}\label{ssec:chiralanom}

Although the whole action in eq.~\eqref{eq:ztilde4} is not U(1) gauge
invariant because of the coupling-dependent mass term in
\eqref{eq:ztilde4}, the action in the path integral which yields the
determinant of the Dirac operator,
\eq{\Det \Dir = \int \cD \psib \cD \psi e^{i\int \extd^2 x \psib \Dir
    \psi}\, ,}{eq:diracdet}
is invariant under $U(1)$ gauge transformations (both in Minkowski and
Euclidean space)
\eq{\de_\La \psi = - i\La \psi\,,\quad \de_\La\psib = i\psib \La\,,\quad \de_\La A_\mu = -\la^{-1}(\partial_\mu \La)\,.}{eq:U1gaugetransf}
Here the vector potential $A_\mu$ is treated as a background field,
and $\La$ is a real gauge potential. It is also invariant under global
chiral transformations,
\eq{\tilde{\de}_\varphi \psi = -i \varphi \ga_\ast \psi\,,\quad \tilde{\de}_\varphi\psib = -i \psib\varphi\ga_\ast\,,\quad (\varphi \in \mathbb{R}) \,.}{eq:chirtrans1}
In Euclidean space this transformation rule still applies, but with
the conventions chosen in App.~\ref{conventions} $\ga_\ast$ becomes
anti-hermitian, in contrast to the Minkowski case where it is
hermitian.  Thus in the Euclidean case a hermitian representation of
the chiral symmetry group $U(1)$ rather than a unitary one is used. To
gauge this chiral symmetry fermions have to be coupled to an axial
vector $A_\mu^5$ (the gauge potential for the chiral symmetry)
transforming as
\eq{\tilde{\de}_\varphi A_\mu^5 = \left\lbrace
\begin{array}{ll}
- \la^{-1} \partial_\mu \varphi & \quad\mathrm{(M)} \\
\ \,i \la^{-1} \partial_\mu \varphi & \quad\mathrm{(E)}
\end{array}\right.\,,
}{eq:A5trans}
for Minkowski (M) and Euclidean (E) space, respectively.  The vector
potential $A_\mu$ is invariant under chiral transformations, as is the
axial vector under gauge transformations.  The locally chiral and
gauge invariant action then reads
\beqa
S & = & (\psi,\Dir\psi) = \int \extd^2 x \sqrt{\mp g} \, \psib \Dir  \psi\label{eq:locchirinvaction}\\\label{eq:diracops}
\Dir & = & \left\lbrace
    \begin{array}{ll}
      iE_a^\mu\ga^a(\nabla_\mu -i\la A_\mu -i\la \ga_\ast A_\mu^5) & \quad\mathrm{(M)} \\
      iE_a^\mu\ga^a(\nabla_\mu -i\la A_\mu +\la \ga_\ast A_\mu^5)  & \quad\mathrm{(E)}
    \end{array}
\right.\,,
\eeqa
with $\psib = \psi^\dagger \ga^0$ for Minkowski space and $\psib =
\psi^\dagger$ for Euclidean space.  The Euclidean Dirac operator and
its square transform as
\beqa
\de_\La \Dir &=& i [\Dir,\La]\,,\qquad \tilde{\de}_\varphi\Dir = i \{ \varphi\ga_\ast,\Dir \} \,,\label{eq:diracchiral} \\
\de_\La D &=& i [D,\La]\,,\qquad \tilde{\de}_\varphi D = i\{ \varphi \ga_\ast,D\} + 2i\varphi \Dir\ga_\ast \Dir\,.\label{eq:laplacechiral}
\eeqa
Because of the vanishing variation eq.~\eqref{eq:zetavar} of the zeta
function,
\eq{\tilde{\de}_\varphi \zeta(s,D) = -is \Tr \left( [D,\La] D^{-s-1} \right) = i\Tr\left( [D^{-s},\La] \right) = 0\,,}{eq:zetavar1}
U(1) gauge invariance is retained on quantum level. The zeta function
is not invariant under chiral transformations,
\eq{\tilde{\de}_\varphi \zeta(s,D) = - 4is\Tr\left( \varphi \ga_\ast D^{-s} \right) = -4i s \zeta(s,\varphi\ga_\ast,D)\,,}{eq:zetavar2}
but yields a non-vanishing chiral anomaly (for $A_\mu^5 = 0$)
\beqa\nonumber
\cA(\varphi) = \tilde{\de}_\varphi W^{\rm ren} & = & -2i \zeta(0,\varphi\ga_\ast,D) \\\nonumber
& = & -2i a_2(\varphi\ga_\ast,D) \\\nonumber
& = & \frac{1}{2\pi i} \int \extd^2 x \sqrt{g}\,\varphi\, \tr_{\complexc^2}\left[\ga_\ast\left(\frac{R}{6} + E \right)\right]\\\nonumber
& = & - \frac{\la}{2\pi}\int \extd^2 x \sqrt{g} \varphi \epsilon^{\mu\nu} F_{\mu\nu}\\
& = & - \frac{\la}{2\pi}\int \extd^2 x \sqrt{g} \,\varphi \ast(\extd A)\,.\label{eq:chiralanomaly}
\eeqa
Decomposing the one-form $A$ uniquely into a gauge and an axial part
\eq{A= \extd \La + \extd^\dagger \ast \varphi\,,}{eq:Hodgedecomposition}
the field strength only depends on the axial part (with
eqs.~\eqref{eq:starquadrat}, \eqref{eq:ddagger}),
\eq{F = \extd A = \extd \extd^\dagger \ast\varphi = - \extd \ast \extd \varphi\,.}{eq:Fdecompose}
With eq.~\eqref{eq:Laplacian} the chiral anomaly becomes (now in
Minkowski space)
\beqa
\tilde{\de}_{\de\varphi} W  & = & \frac{\la}{2\pi}\int \extd^2 x \sqrt{-g} \,\de\varphi\,\Delta \varphi\\\nonumber
                            & = & W_{\rm WZ}[\varphi+\de\varphi] - W_{\rm WZ}[\varphi]\,,\label{eq:chiralanomaly2}
\eeqa
with $W$ being the Wess-Zumino action \cite{Wess:1971yu}
\eq{W_{\rm WZ} = \frac{\la}{4\pi} \int \extd^2 x \sqrt{-g} (\ast F)\frac{1}{\Delta}(\ast F)\,. }{eq:WessZumino}
The full one-loop effective action thus comprises the Polyakov
part eq.~\eqref{eq:Polyakovaction} and the Wess-Zumino part
eq.~\eqref{eq:WessZumino},
\eq{W_{\rm 1loop} = -\ln \Det \Dir = W_{\rm Pol} + W_{\rm WZ}\,.}{eq:W1result}
What remains to be evaluated is the path integral over the vector
potential in eq.~\eqref{eq:ztilde4}, now with the highly non-local
integrand eq.~\eqref{eq:WessZumino}. The nature of the application
should thus decide whether it is favorable to use this form, or to
treat the Thirring term directly as an interaction vertex. On the
other hand it may well be that the Wess-Zumino action becomes local in
some gauges, as happens for the Polyakov action in conformal gauge.
Otherwise auxiliary scalar fields can be introduced to bring the
Polyakov and Wess-Zumino action to local form,
\beqa\label{eq:WPollocal}
W_{\rm Pol} & = & \frac{1}{48\pi}\int \extd^2 x \sqrt{-g} \left[ \frac{1}{2} (\nabla \Phi)^2 + \Phi R\right] \\\label{eq:WZlocal}
W_{\rm WZ}  & = & \frac{\la}{2\pi}\int \extd^2 x \sqrt{-g} \left[ \frac{1}{2} (\nabla Y)^2 + Y (\ast F)\right]\,.
\eeqa
These expressions coincide with the ones in
\cite{Nojiri:1992st,Ori:2001xc}.

Rewriting the Thirring term as in \eqref{eq:rewrThirring} is actually
the application of the bosonization prescription found by Coleman,
Jackiw and Susskind \cite{Coleman:1975pw,Coleman:1974bu} for the
Schwinger model in flat Minkowski space. Transforming to a gauge with
$\La = 0$ in eq.~\eqref{eq:Hodgedecomposition} leads to the
identification $\epsilon^{\mu\nu}
\partial_\mu \varphi \propto \chib \ga^\mu \chi$. The treatment in
this section thus shows that bosonization is also applicable within
the framework of quantized dilaton gravity in two dimensions even
after the geometric sector is quantized nonperturbatively.

\section{Lowest Order Vertices}\label{sec:vertices}

Because the fermions appear as bilinears in the effective action
eqs.~\eqref{eq:Leff2} and \eqref{eq:ambiguous}, the lowest order
interaction vertices are non-local four-point vertices, and all
vertices have an even number of outer legs. In the following only
massless fermions without self interactions ($g(\chib\chi)=0$) are
considered, thus the vertices are solely generated by gravitational
interaction.

\subsection{Effective Geometry and the Virtual Black Hole}\label{ssec:effgeom}

The standard way for extracting the form of interaction vertices would
be to expand the effective action eqs.~\eqref{eq:Leff2} and
\eqref{eq:ambiguous} up to fourth order in the matter fields and read
off the vertex as the corresponding coefficient. This is very
cumbersome because of nested integrals coming from the inverse
derivatives, whose ranges of integration have to be treated properly.
We therefore adopt a strategy introduced in \cite{Kummer:1998zs} and
directly solve the equations of motion for the canonical pair
$(q^i,p_i)$ following from the action eq.~\eqref{eq:Leff1} (together
with the first order gravity constraints eqs.~\eqref{G1g}-\eqref{G3g}
and for vanishing sources $j_i=J^i = 0$),
\beqa\label{eq:vertexp1}
\partial_0 p_1 & = & p_2 \\\label{eq:vertexp2}
\partial_0 p_2 & = & -2 f(p_1) \Phi_0 \\\label{eq:vertexp3}
\partial_0 p_3 & = & -U(p_1) p_2 p_3 - V(p_1) + 2 f(p_1) \Phi_1
\eeqa

\beqa
\label{eq:vertexq1}
\partial_0 q^1 & = & q^3(U'(p_1)p_2p_3 + V'(p_1)) + 2f'(p_1)\big[ q^2 \Phi_0 - q^3 \Phi_1 - \Phi_2 \big] \\\label{eq:vertexq2}
\partial_0 q^2 & = & -q^1 + q^3 p_3 U(p_1)\\\label{eq:vertexq3}
\partial_0 q^3 & = & q^3p_2 U(p_1)\,,
\eeqa
with matter contributions
\eq{\Phi_0 = \frac{i}{2\sqrt{2}}(\chid_1 \lrpd{0} \chi_1)\,,\quad \Phi_1 = \frac{i}{2\sqrt{2}}(\chid_0 \lrpd{0} \chi_0)\,,\quad \Phi_2 = \frac{i}{2\sqrt{2}}(\chid_1 \lrpd{1} \chi_1)}{eq:vertexmatter}
localized at a point $y$ in space-time, i.e.
\eq{\Phi_i = c_i \de^{(2)}(x-y)\,,\ i=0,1,2\,.}{eq:localize}
Such a matter configuration is clearly off-shell, i.e. it does not
fulfill the Dirac equation \eqref{eq:diracchi0}-\eqref{eq:diracchi1}.
Because of the structure of the interaction terms in the effective
action eqs.~\eqref{eq:Leff2} and \eqref{eq:ambiguous}, one can expect
to find three four-point vertices
\beq\nonumber
V^{(4)} = \int_x \int_z \Big[ V_a(x,z) \Phi_0(x)\Phi_0(z)  +
V_b(x,z)\Phi_0(x) \Phi_2(z) + V_c(x,z)\Phi_0(x) \Phi_1(z) \Big]\,.
\eeq
After substituting the solutions of
eqs.~\eqref{eq:vertexp1}-\eqref{eq:vertexq3} back into the interaction
terms of eq.~\eqref{eq:Leff1} and expanding up to second order in the
$c_i$ the explicit form of the vertices is found. In order to be able
to substitute the coefficients $c_i$ by the corresponding bilinears
\eqref{eq:vertexmatter} at the end, one should carefully keep track of
the (because of non-locality of the interaction) different space-time
points where the external legs are situated at.


The solutions of eq.~\eqref{eq:vertexp1}-\eqref{eq:vertexq3} will have
a delta peak in $x^1$ direction and the following continuity
properties: $p_2$, $p_3$, $q^1$ will jump at $y^0$ because of the
delta distributions on the right hand side, while the other variables
will be continuous. To solve these equations unambiguously, a
prescription for the behaviour of the vacuum solutions far away from
the localization point is needed.  To fix the six integration
constants one requires \cite{Grumiller:2002dm} in the asymptotic
region $x^0>y^0$
\begin{itemize}
\item $p_1\big|_{x^0>y^0} = x^0$ and $p_2\big|_{x^0>y^0} = 1$, i.e.
  identification the dilaton with the $x^0$-coordinate. This
  corresponds to the fundamental patch used in
  sec.~\ref{sec:classsol}, and can always be reached from the general
  solutions, eq.~\eqref{eq:p1gvs} and \eqref{eq:p2gvs} below, by a
  residual gauge transformation eq.~\eqref{eq:residuallarge} with
  $\ga(x^1) = - \ln c(x^1)$ and $g(x^1) = - d(x^1)/c(x^1)$ for
  $c(x^1)\neq 0$.

\item $\cC^{(g)}\big|_{x^0>y^0} = \cC_\infty$ fixes the integration
  constant in $p_3$ (see below).

\item $q_3\big|_{x^0>y^0} = e^{Q(x^0)}$ then solves
  eq.~\eqref{eq:vertexq3} and defines the asymptotic unit of length.

\item The remaining two integration constants enter
  $q^2\big|_{x^0>y^0}$, which is the solution of a second order
  partial differential equation derived below, are called $m_\infty$
  and $a_\infty$ because for spherical reduced gravity they correspond
  to the Schwarzschild mass and the Rindler acceleration 
  of the asymptotic background.

\end{itemize}
The general solutions for eqs.~\eqref{eq:vertexp1}-\eqref{eq:vertexp3}
and eq.~\eqref{eq:vertexq3} in the vacuum regions $x^0\neq y^0$ are
\beqa\label{eq:p1gvs}
p_1 &=& c(x^1) x^0 + d(x^1) \\\label{eq:p2gvs}
p_2 &=& c(x^1) \\\label{eq:p3gvs}
p_3 &=& e^{-Q(p_1)}\left[ C(x^1) - \frac{w(p_1)}{p_2}\right] \\\label{eq:q3gvs}
q^3 &=& e(x^1) e^{Q(p_1)}\,.
\eeqa
For finding the vertices the solution for $q^1$ will not be necessary
and, as only the exterior derivative of the spin connection enters the
Ricci scalar
\eq{R = 2\ast \extd \om = -\frac{2}{q^3} \partial_0 q^1\,,}{eq:REFgauge}
the scalar curvature can be read off directly from the right hand side
of eq.~\eqref{eq:vertexq1}. Another differentiation of
eq.~\eqref{eq:vertexq2} with respect to $x^0$ and use of the other
equations of motion yields a second order differential equation for
$q^2$,
\eq{\partial_0^2 q^2 = - e(x^1)w''(p_1) - 2 f'(p_1)\big[ q^2 \Phi_0 - q^3 \Phi_1 - \Phi_2 \big] + 2 f(p_1) q^3 U(p_1) \Phi_1\,,}{eq:vertexq22}
which is solved in the vacuum regions by
\eq{q^2 = m(x^1) + a(x^1) p_1 - e(x^1)\frac{w(p_1)}{p_2^2}\,.}{eq:q2gvs}
Adjusting the integration constants in the region $x^0 > y^0$ as
listed above and patching the solutions at $x^0 = y^0$ according to
their continuity properties ($p_2$, $p_3$, $\partial_0 q^2$ jumping,
$p_1$, $q^3$, $q^2$ continuous) then fixes the integration constants
in the region $x^0<y^0$, yielding up to linear order in the $c^i$
\beqa\label{eq:p1effgeom}
p_1 & = & x^0 + 2 f(y^0) (x^0-y^0)h_0 \\\label{eq:p2effgeom}
p_2 & = & 1 + 2 f(y^0) h_0 \\\nonumber
p_3 & = & e^{-Q(p_1)}\Big[ \cC_\infty - w(p_1) + 2 f(y^0)h_0 (w(x^0) - w(y^0))\\\label{eq:p3effgeom}
    &   & \hspace{4cm}- 2 f(y^0) e^{Q(y^0)} h_1 \Big]\\\nonumber
q^2 & = & m_\infty + a_\infty x^0 - w(p_1) + 2 h_0 \Big[ 2 f(y^0)(w(x^0)-w(y^0))\\\nonumber
    &   & \hspace{1,25cm}+ \big[ (m_\infty + a_\infty y^0) f'(y^0) - (f w)'|_{y^0} \big] (x^0-y^0)\Big]\\
    &   & \hspace{1,3cm}-2 (f e^Q)'|_{y^0} (x^0-y^0) h_1 -2 f'(y^0)(x^0 - y^0) h_2 \label{eq:q2effgeom}\\
q^3 & = & e^{Q(p_1)}\label{eq:q3effgeom}\,,
\eeqa
with $h_i = c_i \theta (y^0 - x^0) \de(x^1 - y^1)$. The three
constants $m_\infty$, $\cC_\infty$ and $a_\infty$ are not independent
from each other because the solutions in the asymptotic region
$x^0>y^0$ have to fulfill the geometric secondary constraints
eq.~\eqref{G1g}-\eqref{G3g} to describe a consistent background. The
Lorentz constraint eq.~\eqref{G1g} then requires
\eq{\cC_\infty = m_\infty\,,\quad a_\infty = 0\,.}{eq:Cma}
Eq.~\eqref{G2g} even yields a vacuum solution
for the spin connection,
\eq{q^1 = \frac{q^3}{p_2} \cV(p_1;p_2p_3)\,,}{eq:q1}
and eq.~\eqref{G3g} is identically fulfilled. With these solutions the
geometric part of the conserved quantity
eq.~\eqref{eq:conservedquantity}
\eq{\cC^{(g)} = m_\infty (1 + 2 f(y^0) h_0) - 2 f(y^0) w(y^0) h_0 -
  2 f(y^0) e^{Q(y^0)} h_1}{eq:Cgeffgeom}
and the scalar curvature eq.~\eqref{eq:REFgauge}
\eq{R = -2 (U'(p_1)p_2p_3 + V'(p_1)) - 4f'(p_1) e^{-Q(p_1)}\big[q^2\Phi_0 - q^3\Phi_1 - \Phi_2\big]}{eq:REFsol}
both jump at $x^0 = y^0$. For non-minimal coupling the Ricci scalar
even has a delta-like singularity at the point $y$, whereas for
minimal coupling and $U'=0$ it becomes continuous. The effective line
element
\eq{(\extd s)² = 2 e^+ \otimes e^- = 2q^3\extd x^1 \otimes \big[\extd x^0 + q^2 \extd x^1 \big]}{eq:effds21}
takes Eddington-Finkelstein form
\eq{(\extd s)^2 = 2 \extd r \extd u + K(r,u,r_0,u_0) \extd u^2}{eq:effds22}
by introducing new coordinates $\extd r = b q^3(x^0) \extd x^0\,,\
b>0$, $\extd u = b^{-1} \extd x^1$ ($b$ is a real scale factor). The
quantity
\begin{multline}
K(r,u,r_0,u_0) = K_\infty\big[ 1 + 2f(y^0)U(x^0)(x^0 - y^0) h_0 \big] - 4 b^2 e^{Q(x^0)} (x^0-y^0) \times\\
                  \times \Big[ \left(f(y^0) (w'(x^0) + w'(y^0)) + f'(y^0) (w(y^0) - m_\infty)\right)h_0 + (f e^{Q})'|_{y^0} h_1 + f'(y^0) h_2 \Big]\\ 
                  + 8b^2e^{Q(x^0)}f(y^0) (w(x^0)-w(y^0)) h_0\label{eq:effkillingnorm}
\end{multline}
with
\eq{K_\infty = 2b^2 e^{Q(x^0)} (m_\infty - w(x^0))}{eq:asymptkn}
is the analogue of the Killing norm of the vacuum solutions,
$K_\infty$. It is continuous at $x^0 = y^0$, as is the whole line
element.  The Ricci scalar \eqref{eq:REFsol} following from the Hodge
dual of the spin connection is however not the same as the Ricci
scalar computed from the line element
\eqref{eq:effds22}-\eqref{eq:asymptkn}, because it depends on the (for
$U(X)\neq 0$) non-vanishing torsion part of the spin connection (cf.
eq.~\eqref{eq:solomT}). The ``torsion free'' Ricci scalar in the
asymptotic region $x^0 > y^0$ reads
\eq{\tilde{R} = \partial_r^2 K_\infty(r) = 2(V'(p_1) + p_2p_3U'(p_1)) - 4 e^{-Q(p_1)} w''(p_1)}{eq:Rtildeas}
and coincides in the absence of torsion ($U(X)=0$) with
\eqref{eq:REFsol}.

For minimally coupled spherical reduced gravity (cf. the first model
in table~\ref{tab:1}; $f(X) = \ka$) with $\la = 1/2\,, b= 1/(2\la)=1$
and $m_\infty = 0$ eq.~\eqref{eq:effkillingnorm} contains a Rindler
term ($\propto r$) and a Schwarzschild term ($\propto 1/r$),
\beqa\label{eq:effkillingnormSRG}
K(r,u,r_0,u_0) & = & K_\infty - 4 \ka h_0 - \frac{2M}{r} + a r \\
M(r,u,r_0,u_0) & = & -\frac{3}{2} \ka r_0 h_0 + \frac{4 \ka}{r_0} h_1 \label{eq:SRGeffmass}\\
a(r,u,r_0,u_0) & = & +\frac{\ka}{r_0}h_0 + \frac{8\ka}{r_0^3}h_1\label{eq:SRGeffaccel}\,.
\eeqa
The dilaton coupling function $f(y^0)$ enters
\eqref{eq:effkillingnorm} only evaluated at the localization point $y$
and thus the structure of the effective line element
\eqref{eq:effds22} does not change, showing that the virtual black
hole is a generic feature for both minimal and non-minimal coupling.

As the asymptotic region $x^0\rightarrow \infty$ in this example is
Minkowski space ($m_\infty = 0$, $K_\infty = 1$), the occurrence of
such an effective geometry is due to the scattering process, and in
particular the occurrence of the Schwarzschild term is interpreted as
the formation of a virtual black hole as an intermediary scattering
state \cite{Grumiller:2000ah}.\newpage
\begin{wrapfigure}{r}{3cm}
\includegraphics[height=6cm]{virtualBH.epsi}
\caption{VBH}
\label{fig:CP}
\end{wrapfigure}
Fig.~\ref{fig:CP} shows the
Carter-Penrose diagram for the resulting space-time. The non-trivial
geometry is located on a light-like cut $u = u_0$ extending from $r=0$
up to a maximal radius $r \le r_0$. The geometry is non-local in the
sense that it depends on two space-time coordinates $(r_0,u_0)$ and
$(r,u)$. The rest of space-time is flat Minkowski space or, in the
general case, the background given by the line element
eq.~\eqref{eq:effds22} with the asymptotic Killing norm
eq.~\eqref{eq:asymptkn}. As a word of warning this intermediary
geometry should not be over-interpreted. It is off-shell, as the
localized matter contributions are, and it is as much virtual as
virtual particles in ordinary perturbative quantum field theory are.

\subsection{Four-Point Vertices}

Inserting the solutions eqs.~\eqref{eq:p1effgeom}-\eqref{eq:q3effgeom}
and \eqref{eq:Cma} into the interaction terms of eq.~\eqref{eq:Leff1}
(cf. also \eqref{eq:vertexmatter} and \eqref{eq:localize}),
\eq{S_{\rm int} = -2 \int\limits_x f(p_1)\big[ q^2 \Phi_0 - q^3 \Phi_1 - \Phi_2 \big]\,,}{eq:Leff1ww}
expanding up to first order in the $c^i$, replacing these coefficients
with $\Phi_i(y)$ and integrating over $y$ yields three four-point
vertices, namely the symmetric one
\begin{multline}
V_a = -4 \int\limits_x \int\limits_y \Phi_0(x) \Phi_0(y) \theta(y^0-x^0) \de(x^1 - y^1) f(x^0) f(y^0) \times \\
\times \Big[ 2(w(x^0)-w(y^0)) - (x^0-y^0)(w'(x^0) + w'(y^0)) \\
- (x^0 - y^0) \Big(\frac{f'(x^0)}{f(x^0)}(w(x^0) - m_\infty) + \frac{f'(y^0)}{f(y^0)}(w(y^0) - m_\infty)\Big)\Big]
\label{eq:Va}
\end{multline}
and two non-symmetric ones
\beqa
V_b & = & -4\int\limits_x \int\limits_y \Phi_0(x) \Phi_2(y) \de(x^1 - y^1) |x^0-y^0| f(x^0) f'(y^0)\,.\label{eq:Vc}\\
V_c & = & -4\int\limits_x \int\limits_y \Phi_0(x) \Phi_1(y) \de(x^1 - y^1) |x^0-y^0| f(x^0) (f e^Q)'|_{y^0}\label{eq:Vb}
\eeqa
Interestingly, the vertices $V_a$ and $V_b$ are the same as for a real
scalar field \cite{Grumiller:2002dm}, and only $V_c$ is new. They
share some properties with the scalar case, namely:
\begin{enumerate}
\item They are local in the coordinate $x^1$, and non-local in $x^0$.
\item They vanish in the local limit ($x^0\rightarrow y^0$) and $V_b$ vanishes for minimal coupling.
\item They respect a $\mathbb{Z}_2$ symmetry $f(X)\mapsto - f(X)$.
\item The symmetric vertex depends only on the conformally invariant
  combination $w(X)$ and the asymptotic value $m_\infty$ of the
  geometric part of the conserved quantity
  \eqref{eq:conservedquantity}. $V_b$ is independent of $U$, $V$ and
  $m_\infty$. Thus if $m_\infty$ is fixed in all conformal frames,
  which is ensured by the conformal transformation properties listed
  in sec.~\ref{sec:confEF}, both vertices are conformally invariant.
\end{enumerate}
The new vertex $V_c$ is not conformally invariant because it contains
$U(y^0)$, which is mapped to $\tilde{U}(y^0)$ under a change of the
conformal frame. Still, conformal invariance of the four-fermi
scattering matrix elements is retained at tree-level, because for
on-shell matter configurations (cf.
eqs.~\eqref{eq:solChi0}-\eqref{eq:vertexmatteronshell} below) the
fermion bilinear $\Phi_1$ and thus the vertex $V_c$ vanishes. As all
vertices containing $\Phi_1$ are generated by the last term in
eq.~\eqref{eq:ambiguous} it can be concluded that all tree-level
Feynman diagrams containing pairs of outer $\chi_0$-legs attached to
the same vertex must vanish, and all non-vanishing Feynman diagrams
containing outer $\chi_0$-legs also have to contain internal
$\chi_0$-propagators. Whether these diagrams also vanish or whether
they yield conformally invariant S-matrix elements is not guaranteed
{\it a priori}, but tree-level conformal invariance of the S-matrix is
expected to hold because of the nonperturbative result obtained in
sec.~\ref{sec:confEF}. At one-loop level, however, conformal
invariance is broken by the conformal anomaly, cf.
sec.~\ref{ssec:confanom}.

Thus as in the scalar case, where for minimal coupling the
four-particle scattering amplitude for spherical reduced gravity (cf.
the first model in table~\ref{tab:1}) turned out to be either zero or
infinity \cite{Grumiller:2000ah} depending on which boundary
conditions at the origin where chosen for the asymptotic states, only
the symmetric vertex $V_a$ will contribute at tree-level.

\section{Asymptotic Matter States}\label{sec:asmat}


The metric obtained from eq.~\eqref{eq:effds21} (and
eqs.~\eqref{eq:q2effgeom} and \eqref{eq:q3effgeom} with $a_\infty =
0$) in the region $x^0 \rightarrow \infty$ determines the asymptotic
states for calculation of the scattering matrix. If they form a
complete and (in an appropriate sense) normalizable set, an asymptotic
Fock space can be constructed. In the Dirac equation obtained from
eq.~\eqref{eq:Lkin} with the Zweibein in Eddington-Finkelstein gauge
\eqref{eq:EFgauge} and with
\eq{q^2(x^0) = m_\infty - w(x^0)\,,\quad q^3(x^0) = e^{Q(x^0)}\,,\quad f(p_1) = f(x^0)}{eq:q2q3asympt}
the spinor components decouple for massless fermions,
\beqa\label{eq:diracchi0}
\partial_0 \chi_0 & = & -\frac{1}{2}\left(\frac{f'(x^0)}{f(x^0)} + U(x^0)\right) \chi_0 \\\label{eq:diracchi1}
(\partial_1 - q^2(x^0) \partial_0) \chi_1 & = & \frac{1}{2}\left(\frac{f'(x^0)}{f(x^0)} q^2(x^0) +  {q^2}'(x^0) \right)\chi_1\,.
\eeqa
These equations are just the Dirac equation $\Dir \chi = 0$ (cf.
eq.~\eqref{eq:diracop}) in components, with a spin connection
\beqa
\om_0 & = & -U(x^0) - \frac{f'(x^0)}{f(x^0)} \\
\om_1 & = & -U(x^0) q^2(x^0) - {q^2}'(x^0) -2 \frac{f'(x^0)}{f(x^0)}q^2(x^0)\,.
\eeqa
The equation for $\chi_1$ is conformally invariant (with conformal
weights as in table~\ref{tab:conf}), but the one for $\chi_0$ contains
$U(X)$ and thus transforms while changing between conformal frames. To
solve the second equation one introduces new coordinates ($C$, $C'$
are constants)
\beqa
v(x^0,x^1) & = & x^1 - \int\limits^{x^0} \frac{\extd z}{q^2(z)} + C  \label{eq:vcoordinate} \\
w(x^0,x^1) & = & x^1 + \int\limits^{x^0} \frac{\extd z}{q^2(z)} + C'
\label{eq:wcoordinate} \eeqa
in patches where the integrals are defined. In this way both
coordinates are locally orthogonal to each other, $\partial_v w =
\partial_w v = 0$. With the Ansatz
\eq{\chi_i = \cR_i e^{i\phi_i}\,;\quad \cR_i,\phi_i \ {\rm real}}{eq:chiiansatz}
one finds that $\cR_0$ has to fulfill eq.~\eqref{eq:diracchi0},
$\cR_1$ has to obey
\eq{\partial_v \cR_1 = \frac{1}{4}\left(\frac{f'(x^0)}{f(x^0)} q^2(x^0) + {q^2}'(x^0) \right)\cR_1\Big|_{x^0=x^0(v,x^1)}\,,}{eq:cR1v}
and the phases $\phi_i$ obey
\eq{\partial_0 \phi_0 = 0\quad {\rm and}\quad  \partial_v\phi_1 = 0\,.}{eq:phases}
Most notably, solutions can be obtained explicitly even for general
non-minimal coupling,
\beqa\label{eq:solChi0}
\chi_0(x^0,x^1) & = & \frac{F(x^1)}{\sqrt{f(x^0)}}\, \exp \Big[{i\phi_0(x^1)-\frac{Q(x^0)}{2}}\Big]\\
\chi_1(x^0,x^1) & = & \tilde{F}(w)e^{i\phi_1(w)}\times\nonumber\\ 
                &   & \hspace{-2cm}\times \exp\left[ \frac{1}{4} \int\limits^{v(x^0,x^1)} \extd v' \left( \frac{f'(x^0)}{f(x^0)}q^2(x^0)  + {q^2}'(x^0) \right)\Big|_{x^0 = x^0(v',x^1)}\right]\,,\label{eq:solchi1}
\eeqa
with $F(x^1)$ and $\tilde{F}(w)$ being arbitrary but real functions.
The conformal transformation properties of these solutions are as
expected in sec.~\ref{sec:confEF}. $\chi_0$ includes a factor
$e^{-Q(x^0)/2}$ and thus transforms with weight $-1$, while $\chi_1$
only depends on the conformally invariant combination $e^{Q(x^0)}
V(x^0)$ and the mass $m_\infty$ of the asymptotic space-time.  Thus if
$m_\infty$ is fixed for all conformal frames, the solution for
$\chi_1$ is conformally invariant. The fermion bilinears
\eqref{eq:vertexmatter} which enter the tree-level scattering vertices
(e.g. the four-fermi vertices \eqref{eq:Va}-\eqref{eq:Vc}) read
on-shell
\eq{\Phi_0(x) = - \frac{|\chi_1|²}{\sqrt{2}}\frac{\phi_1'(w)}{q^2(x^0)} = \frac{\Phi_2(x)}{q^2(x^0)}\,,\quad \Phi_1(x) = 0\,. }{eq:vertexmatteronshell}
As noted at the end of the last section, the vanishing of $\Phi_1$
implies conformal invariance of the tree-level scattering matrix
elements for four-fermion scattering as well as vanishing of a certain
class of tree-level Feynman diagrams.


\chapter{Conclusions and Possible Further Developments}\label{ch:conclusions}

\section{Summary and Conclusions}

The goal of this thesis was to analyze the classical structure and to
quantize two-dimensional dilaton gravity models in the first order
formulation coupled to Dirac fermions using the Feynman path integral
approach. Chapter~\ref{ch:intro} contained a short motivation for
considering gravity in two dimensions in general and first order
gravity with fermions in particular, together with some historical
remarks.

\subsection{Classical Analysis}

In chapter~\ref{ch:classical}, after introducing the action describing
the system under consideration, the equivalence between the first
order formulation and the second order dilaton gravity action on the
classical and quantum level was demonstrated, for which it was crucial
that fermions in two dimensions do not directly couple to the spin
connection and the Lagrange multipliers for torsion. After recalling
how to obtain all classical solutions for the matterless theory or
even for special matter configurations, namely (anti)chiral fermions
and (anti)self-dual spinors, the constraints of the system were
analyzed. Three primary first class constraints and four well-known
primary second class constraints relating the spinor to its canonical
momentum, as well as three secondary first class constraints
generating local Lorentz transformations and infinitesimal
diffeomorphisms were found. The Hamiltonian turned out to be fully
constrained, as expected for a generally covariant theory.

The algebra of the secondary first class constraints,
eqs.~\eqref{GiGi}-\eqref{G2G3}, is the main result of this analysis
and should be compared to the known cases of first order gravity
without matter \cite{Kummer:1996hy} and with a real scalar field
\cite{Kummer:1998zs,Grumiller:2001ea}. As for scalar matter, only the
bracket between the two diffeomorphism generators \eqref{G2G3} is
changed in comparison to the matterless theory, but in the very
specific way that the matter contribution is proportional to the local
Lorentz generator. A new feature is that not only for minimal coupling
but also for certain non-minimal couplings ($f \propto h$) the
constraint algebra is the same as in the matterless case, namely if
the fermions have non-vanishing mass but otherwise do not
self-interact ($g(\chib\chi) = m \chib\chi$).  Also the kinetic term
of the fermions does not contribute at all, whereas in the scalar case
\cite{Grumiller:2001ea} the additional structure function was just the
kinetic term for the real boson. Furthermore, the algebra loses its
property to be a finite W-algebra.

It is remarkable that the constraint algebra \eqref{GiGi}-\eqref{G2G3}
closes without any spatial derivatives of delta functions and in this
sense resembles rather an ordinary gauge theory or Ashtekar's approach
to gravity \cite{Ashtekar:1986yd,Ashtekar:1987gu}, while the linear
combinations \eqref{eq:newgen} correspond to the Hamiltonian ($H_0$)
and diffeomorphism constraint ($H_1$) of the ADM approach
\cite{Arnowitt:1962} and fulfill the Virasoro algebra which closes
with derivatives of delta functionals.

The treatment of boundaries for the matterless theory was also briefly
discussed. It turns out that without any additional boundary terms in
the first order gravity action \eqref{eq:FOG} most of the constraint
algebra is unchanged, but eq.~\eqref{G2G3} receives a boundary
contribution \eqref{eq:G2G3bdry} which vanishes for $(A)dS_2$ ground
state models.\footnote{Maybe one should further investigate whether
  there are any connections of this point to the topic of ${\rm
    AdS}_2$/${\rm CFT}_1$ correspondence.}

At the end of chapter~\ref{ch:classical} the BRST charge was
constructed and Eddington-Finkelstein gauge \eqref{eq:EFgauge} on the
manifold and light cone gauge in tangent space was fixed. The
homological perturbation theory terminates at Yang-Mills level, which
can be inferred from the Poisson-$\sigma$ model structure
\cite{Schaller:1994es} of the matterless theory but is a non-trivial
fact in the presence of matter when the theory does not admit a
Poisson-$\sigma$ model formulation. In general one would expect the
presence of higher order ghost terms. The choice \eqref{eq:EFgauge}
does not fix the gauge completely but allows for residual gauge
transformations \eqref{eq:residuallarge}, namely combined Lorentz
transformations and diffeomorphisms.

\subsection{Nonperturbative Quantization of Geometry}

In chapter~\ref{ch:nonpert}, the main part of this thesis, the path
integral over the ghost sector and the geometric fields which were not
fixed by the gauge choice \eqref{eq:EFgauge} was performed
nonperturbatively. I want to emphasize that this is only possible for
two reasons. First, the constraints \eqref{G1}-\eqref{G3} which appear
in the Hamiltonian density are linear in the fields $q^i =
(\om_1,e_1^-,e_1^+)$, and second, the gauge choice \eqref{eq:EFgauge}
not only further simplifies the structure of the Hamiltonian but also
generates a Faddeev-Popov determinant \eqref{eq:FPdet} which solely
depends on the canonical momenta $p_i = (X,X^+,X^-)$. The integration
over $q^i$ can then be carried out first, leading to the rather simple
set of equations \eqref{eq:p1}-\eqref{eq:p3}, and in the course of
subsequently integrating over the $p_i$ the Faddeev-Popov determinant
then cancels.

Another remarkable point about this procedure is that the integration
over the Zweibeine $e_1^\pm$ is not restricted at all. In
Eddington-Finkelstein gauge \eqref{eq:EFgauge} the volume element is
$\sqrt{-g} = e_1^+$, which classically should be restricted to
positive values. But in order to carry out the nonperturbative path
integration one has to integrate over the whole range of $e_1^+$ to
arrive at the classical equations of motion for the $p_i$
eqs.~\eqref{eq:p1}-\eqref{eq:p3}. Including quantum fluctuations of
the metric corresponding to zero and negative volumes may thus also be
the correct approach for quantizing other gravity theories, including
general relativity.

Furthermore, this quantization of the geometric degrees of freedom is
background independent in the sense that the metric has not been
divided into a fiducial and a fluctuation part.  Although the
homogeneous parts in the solutions of eqs.~\eqref{eq:p1}-\eqref{eq:p3}
(the $\tilde{p}_i$) fix an asymptotic background at the infinitely far
boundary, the geometry in the interior of the manifold is subjected to
the quantum dynamics of the system and the quantization procedure does
not depend on specific properties of special asymptotic backgrounds
but is valid for all choices.

The resulting nonlocal and nonpolynomial effective action
eqs.~\eqref{eq:Leff2} and \eqref{eq:ambiguous} depends on the spinors,
external sources for the geometry and the asymptotic values
$\tilde{p}_i$. It includes three ambiguous terms, which arise as a
homogeneous solution of the (regularized) partial time derivative
$\nabla_0$. They have to be present for several reasons discussed in
sec.~\ref{sec:ambiguous}, namely ensuring quantum triviality of first
order gravity without matter and for agreement with results obtained
in cases where the more ``natural'' order of integrating first the
momenta $p_i$ and then the coordinates $q^i$ (as e.g.  in the
Katanaev-Volovich model, the fourteenth entry in table~\ref{tab:1}) is
possible. At the end of chapter~\ref{ch:nonpert} the resulting
effective action was shown to be conformally invariant at tree-level
in the matter fields.

\subsection{Matter Perturbation Theory and Bosonization}

Chapter~\ref{ch:matter} treats the remaining path integral over the
Dirac fermions in the usual perturbative approach. The fermions were
shown to propagate on an effective background which consistently
includes matter backreactions to arbitrary orders in the matter
fields. At one-loop level both conformal and chiral symmetries develop
anomalies, which where used to calculate the one-loop effective action
consisting of a Polyakov and a Wess-Zumino part.

In that course it was shown that the quantum equivalence between
bosons and fermions in flat two dimensional space-time found by
Coleman, Jackiw and Susskind \cite{Coleman:1975pw,Coleman:1974bu} can
also be applied to the situation when dilaton gravity is quantized
nonperturbatively. The prescription has been used, with the caveat
that it only holds in regions where the space-time curvature is small
compared to the microscopic length scale of the quantum theory, in
\cite{Frolov:2005ps,Frolov:2006is,Thorlacius:2006tf} for semi-classical studies of the
effects of pair production on the global structure of black hole
space-times. It is now clear that this caveat can be dropped at the
expense of the quantum theory becoming more complicated, with a path
integration over the nonlocal Wess-Zumino action to be evaluated.

In the second part of this chapter the lowest order (i.e.
four-particle) tree-level vertices where calculated for massless, not
self-interacting fermions. They turn out to be nonlocal in the
$x^0$-coordinate and related to an intermediary off-shell geometry
which can be thought of as a geometric state forming during the
scattering process. For spherical reduced gravity (cf. the first model
in table~\ref{tab:1}) the line element of this intermediary state
contains a Schwarzschild and a Rindler term which is interpreted as
the formation of a virtual black hole.

Though being a gauge-dependent feature it is noteworthy that two of
the gravitationally induced four-fermion scattering vertices
(\eqref{eq:Va} and \eqref{eq:Vc}) found are the same as for a real,
massless scalar field coupled to first order gravity
\cite{Grumiller:2002dm}. These two vertices are invariant under
conformal transformations. The third vertex, \eqref{eq:Vb}, although
being new compared to the scalar case and not conformally invariant,
was seen to vanish on-shell (cf.  \eqref{eq:vertexmatteronshell}), and
thus for tree-level four-fermi scattering only the symmetric and
non-symmetric vertices \eqref{eq:Va} and \eqref{eq:Vc} contribute. The
corresponding S-matrix elements are conformally invariant because the
vertices $V_a$ and $V_b$ and the asymptotic states \eqref{eq:solchi1}
are. Furthermore, all tree-level Feynman diagrams containing pairs of
outer $\chi_0$-legs attached to the same vertex are found to vanish,
such that scattering of $\chi_0$-particles into $\chi_1$-particles
necessarily involves the exchange of virtual $\chi_0$-excitations,
i.e. the presence of internal $\chi_0$-propagators in the
corresponding Feynman diagrams.

Minimally coupled ($f(X)={\rm const.}$) models with $U(X)=0$ and
$V(X)={\rm const.}$ are found to be scattering trivial, i.e. all
tree-level vertices vanish, as can be seen from eq.~\eqref{eq:Leff2}
and \eqref{eq:ambiguous}. This is a stronger requirement compared to
the scalar case, where the last term in \eqref{eq:ambiguous} was not
present and where it thus was sufficient to require constancy of the
$w' = e^Q V$-term in \eqref{eq:ambiguous} (and minimal coupling). The
CGHS black hole and Rindler ground state models (cf.
table~\ref{tab:1}) could thus exhibit non-trivial fermion scattering
and the class of scattering trivial models will be strongly
restricted.

\section{Possible Extensions}\label{sec:extens}

In \cite{Grumiller:2001ea} a number of interesting ideas have been
listed for the case of first order gravity coupled to a real scalar
field. Most of the remarks made there also apply to dilaton gravity
with fermionic matter and will not all be repeated here. For other
applications specific to fermions cf. also sec.~6.3. in
\cite{Grumiller:2006ja}.

As all needed tools (asymptotic states, vertices) for the evaluation
of the scattering matrix elements for four-fermion scattering are
provided in this thesis, such a calculation is the next natural step.
For a massless real boson coupled to spherical reduced gravity (cf.
the first model in table~\ref{tab:1}) the amplitude obtained in
\cite{Grumiller:2001ea,Fischer:2001vz} has been shown, among other
things, to be unitary and the cross section to be CPT invariant
\cite{Grumiller:2001rg,Grumiller:2004yq}. This is rather remarkable
for two reasons: First, the effective theory obtained after quantizing
geometry is non-local and thus the CPT theorem \cite{Streater:1989vi}
does not apply. Secondly, as argued first in \cite{Hawking:1982dj} and
clarified later in \cite{Ellis:1983jz}, if the information paradox was
really to hold and pure states could evolve into mixed ones through
quantum gravity effects, then quantum mechanics itself should be
modified to incorporate this non-unitary time evolution which in turn
would lead to CPT violation. Reversely, the found CPT invariance of
the S-matrix in the scalar case shows that the time evolution in these
lowest order tree-level scattering processes is unitary.
Investigation of CPT invariance in higher-order tree-level scattering
and for loop calculations could thus shed some light on the
information paradox.  Also, on a real boson field charge conjugation
acts trivially, while parity transformation is respected per
constructionem in a spherical symmetric theory and time reversal can
be achieved by changing from outgoing to ingoing Eddington-Finkelstein
gauge.  Hence investigating the case of Dirac fermions, on which C and
P act nontrivial too, would be more interesting.

Besides that it is clear now how bosonization works on the level of
the path integral \eqref{eq:ztilde4}, the fact that two of the
vertices calculated in sec.~\ref{sec:vertices} are exactly the same as
for a free massless scalar field coupled to first order gravity
\cite{Grumiller:2002dm} and that the contribution of the new vertex
\eqref{eq:Vb} vanishes for a large class of Feynman diagrams is
another indication that bosonization also holds for the tree-level
S-matrix elements. To clarify this issue the S-matrix elements for
fermion scattering in some specific model (e.g. spherical reduced
gravity) should be compared with the ones from the bosonic side of the
correspondence.  Another remarkable fact is that the structure
functions $C_{ij}{}^k$ and in particular $C_{23}{}^1 = -h'(X) \la
(\chib\chi)^2$ of the constraint algebra (cf.~\eqref{G2G3} for
couplings $f(X)=h(X)$ and $g(\chib\chi) = m\chib\chi +
\la(\chib\chi)^2$) is mapped to its counterpart in a free massless
scalar theory \cite{Grumiller:2001ea} by the identification
\eq{S^\pm = i e^{\pm i \varphi} \sqrt{\la} \chib \ga^\pm \chi\,,\quad S^\pm = \ast (\extd S\wedge e^\pm)\,,}{eq:Cijkboson}
with $S(x)$ being the scalar field and $\varphi$ a phase. The
Faddeev-Popov operators \eqref{eq:FPoperator} of both theories are
mapped onto each other, which however does not affect the quantum
theories because the Faddeev-Popov determinant cancels during the
exact path integral quantization of geometry in both cases. The
constraints $G_i$ themselves are however not identified under this
mapping.

The treatment of boundaries in first order gravity coupled to matter
fields is another open problem. As pointed out by Carlip
\cite{Carlip:2004mn,Carlip:2005xy,Carlip:2006fm}, the universality of
black hole entropy could be connected to the presence of a black hole
horizon, which imposes constraints on the physical phase space, and an
underlying symmetry somehow attached to the horizon. The last point
seems to be connected to a conjecture by 't Hooft
\cite{'tHooft:2004ek} that physical degrees of freedom should be
converted into gauge degrees of freedom at a horizon. Indeed, in
\cite{Bergamin:2005pg,Grumiller:2006rc} this was shown to be the case
for first order gravity \eqref{eq:FOG} without additional matter
fields by constructing the reduced phase space. For the case of a
generic boundary one physical degree of freedom was found, whereas for
a horizon boundary the reduced phase space turned out to be empty. It
would be interesting to see whether such a phenomenon also occurs in
the presence of physical matter degrees of freedom.  An indication in
favor of such a phenomenon is the changing rank of the Dirac matrix
\eqref{Calphabeta} at a horizon $e_1^- = 0$. Boundary conditions for
the matter fields can be introduced by means of imposing constraints
at the boundary. On the technical side the Dirac consistency algorithm
is more involved \cite{BGKMV:Boundaries}, with the Dirichlet boundary
constraint for a scalar field generating an infinite tower of
constraints similar to the flat space example considered in
\cite{Sheikh-Jabbari:1999xd}. The Dirac matrix \eqref{Calphabeta} then
contains terms with distributional support at the boundary (i.e.
$\de(x^1_{\rm boundary})$ terms), so that a naive definition of the
inverse matrix and construction of the Dirac bracket does not seem
feasible. On the other hand, as opposed to the situation in flat space
\cite{Sheikh-Jabbari:1999xd}, on a curved space-time one can not
impose the boundary conditions directly in Fourier space. A possible
way out that could lead to a sensible constraint algebra allowing to
judge which gauge symmetries survive in the presence of boundaries may
be to use a regularization of the $\de$-functions at the boundary,
define the Dirac bracket, calculate the constraint algebra and remove
the regularization afterwards, but our efforts in that direction
\cite{BGKMV:Boundaries} so far have not been successful in removing
all divergent contributions with support at the boundary.



\begin{appendix}

\renewcommand{\chaptermark}[1]{
  \markboth{\appendixname\ \thechapter.\ #1}{}}

\chapter{Notational Conventions}
\label{conventions}

Throughout this thesis natural units $\hbar = c = G_N = 1$ are used.
Space-time is assumed to be a Lorentzian manifold with signature
$(+-)$.  Latin indices $a,b$ are local Lorentz indices, while Greek
indices $\mu,\nu$ refer to the manifold. The totally antisymmetric
symbol both in tangent space and on the manifold is defined by
$\epsilon_{01} = \tilde{\epsilon}_{01} = +1$, $\epsilon_{ab} = -
\epsilon_{ba}$ and $\tilde{\epsilon}_{\mu\nu} = -
\tilde{\epsilon}_{\mu\nu}$. The symbols with upper indices are defined
as the usual matrix inverse. In light cone coordinates
\eq{u^\pm = \frac{1}{\sqrt{2}}(u^0 \pm u^1)}{eq:lccoord}
it becomes ${\epsilon^\pm}_\pm = \pm 1$, where local Lorentz indices
are raised with the flat metric $\eta_{ab} = {\rm diag}(+-)$ and
$\eta_{+-} = \eta_{-+} = 1$. Space-time indices are raised and lowered
with the metric $g_{\mu\nu}$.

The components of a p-form are defined via
\beq
\Om_p = \frac{1}{p!} \Om_{\mu_1 \dots \mu_p} \extd x^{\mu_1} \wedge \dots \wedge \extd x^{\mu_1}\,,
\eeq
and the Hodge dual of a p-form (in D dimensions) is defined with the
coefficients
\beq
\ast \Om_p = \Om'_{D-p} = \frac{1}{p!(D-p)!} {\epsilon_{\mu_1\dots\mu_{D-p}}}^{\nu_1\dots \nu_p} \Om_{\nu_1 \dots \nu_p} \extd x^{\mu_1} \wedge \dots \wedge \extd x^{\mu_{D-p}}\,,
\eeq
with the Levi-Civita tensor $\epsilon_{\mu_1\dots \mu_D} = |\det
e_\mu^a| \tilde{\epsilon}_{\mu_1\dots \mu_D}$. For even dimension $D$
and Lorentzian signature this yields
\beq\label{eq:starquadrat} \ast\ast \Om_p = (-1)^{p+1} \Om_p \eeq
and thus
\beq\label{eq:starepsilon}
\ast \epsilon = 1 \,,\quad \ast 1 = - \epsilon
\eeq
for the Hodge star acting on the volume form
\beqa\label{eq:volelem}
\epsilon & = & (e) \extd^2 x \\\label{eq:detmet}
(e)      & = & \det ({e_\mu}^a) = e_0^+ e_1^- - e_1^+ e_0^-\,.
\eeqa
With these conventions the hermitian conjugate of the exterior
derivative \cite{nakaharageometry} reads
\eq{\extd^\dagger = \ast \extd \ast\,.}{eq:ddagger}
The volume form can be written as
\beqa\label{volform1}
\epsilon^{ab} \epsilon & = & e^a \wedge e^b\\\label{volform2}
\epsilon & = & - \frac{1}{2} \epsilon_{ab} e^a \wedge e^b = e^+ \wedge e^-\,.
\eeqa

A Dirac fermion in two dimensions has two complex components, $\chi = (\chi_0, \chi_1)^{\rm T}$. Dirac matrices in two-dimensional Minkowski space are chosen as
\beq\label{eq:diracmatrices}
\begin{array}{ll}
\ga^0 = \ \;\left(
\begin{array}{rr}
	0 & 1 \\
	1 & 0
\end{array}\right)
& \ \ \ 
\ga^1 = \left(
\begin{array}{rr}
	0 & 1 \\
	-1 & 0
\end{array}\right) \\
\ga^+ = \left(
\begin{array}{rr}
	0 & \sqrt{2} \\
	0 & 0
\end{array}\right)
& \ \ \ 
\ga^- = \left(
\begin{array}{rr}
	0 & 0 \\
	\sqrt{2} & 0
\end{array}\right)\,.
\end{array}
\eeq
The analogue of the $\ga^5$ matrix is $\ga_\ast = \ga_0 \ga_1 = {\rm
  diag}(+-)$. They satisfy $\{\ga^a,\ga^b\} = 2 \eta^{ab}$ and
$\{\ga_\ast,\ga^a\} = 0$. For calculations in Euclidean space $\ga^0$
is defined as above, but $\ga^1 = {\rm diag}(+-)$ and $\ga_\ast =
\ga_0\ga_1$, thus satisfying $\{\ga^a,\ga^b\} = 2 \de^{ab}$. The Dirac
matrices in Euclidean space are thus hermitian, $\ga^a =
{\ga^a}^\dagger$, whereas $\ga_\ast$ becomes anti-hermitian.

\clearplaindoublepage

\chapter{Details from Ch. \ref{ch:classical}}\label{app:B}

\section{Poisson Brackets of the Secondary with the Second Class Constraints}\label{app:B1}
To calculate the Dirac brackets, all Poisson brackets of the $G_i$
with the $\Phi_\al$ are needed. They are easily obtained by using the
algebraic properties of the graded Poisson bracket, eqs.
\eqref{eq:gPBsymmetry}-\eqref{eq:gPBLeibnitz}, and read
\beqas
\spoiss{G_1}{\Phi_0'} & = & -\frac{i}{\sqrt{2}} f e_1^+ \chid_0 \;\de(x-x')\\
\spoiss{G_1}{\Phi_2'} & = & -\frac{i}{\sqrt{2}} f e_1^+ \chi_0  \;\de(x-x')\\
\spoiss{G_1}{\Phi_1'} & = & -\frac{i}{\sqrt{2}} f e_1^- \chid_1 \;\de(x-x')\\
\spoiss{G_1}{\Phi_3'} & = & -\frac{i}{\sqrt{2}} f e_1^- \chi_1  \;\de(x-x')\\
\spoiss{G_2}{\Phi_0'} & = & \frac{i}{\sqrt{2}} \left[ f'+Uf \right] X^+ e_1^+ \chid_0\;\de(x-x')\\
                      &   & -\; e_1^+ h g' \chid_1 \;\de(x-x')\\
\spoiss{G_2}{\Phi_2'} & = & \frac{i}{\sqrt{2}} \left[ f'+Uf \right] X^+ e_1^+ \chi_0\;\de(x-x')\\
                      &   & +\; e_1^+ h g' \chi_1  \;\de(x-x')\,.
\eeqas
A little care is needed when integrating by parts derivatives of the
delta distributions. This is most easily done by smearing the fields
with test functions and use of the identity
\beqs
\int dx \varphi(x)\left[ f(x) - f(y) \right]\partial_x\de(x-y) = - \int dx \varphi(x) (\partial_x f(x))\, \delta(x-y)\,,
\eeqs
thus obtaining
\beqas
\spoiss{G_2}{\Phi_1'} & = & \frac{i}{\sqrt{2}} \left[ \chid_1(\om_1 f - X^+ e_1^- f' - X^- e_1^+ U f)
                            + 2 (\partial_x \chid_1)f  + (\partial_x f)\chid_1 \right] \de\\
                      &   & -\; e_1^+ h g' \chid_0 \;\de\\
\spoiss{G_2}{\Phi_3'} & = & \frac{i}{\sqrt{2}} \left[ \chi_1(\om_1 f - X^+ e_1^- f' - X^- e_1^+ U f)
                            + 2 (\partial_x \chi_1)f  + (\partial_x f)\chi_1 \right] \de\\
                      &   & +\; e_1^+ h g' \chi_0 \;\de\\
\spoiss{G_3}{\Phi_0'} & = & \frac{i}{\sqrt{2}} \left[ \chid_0(\om_1 f - X^- e_1^+ f' - X^+ e_1^- U f)
                            - 2 (\partial_x \chid_0)f  - (\partial_x f)\chid_0 \right] \de\\
                      &   & +\; e_1^- h g' \chid_1 \;\de\\\nonumber
\spoiss{G_3}{\Phi_2'} & = & \frac{i}{\sqrt{2}} \left[ \chi_0(\om_1 f - X^- e_1^+ f' - X^+ e_1^- U f)
                            - 2 (\partial_x \chi_0)f  - (\partial_x f)\chi_0 \right] \de\\
                      &   & -\; e_1^- h g' \chi_1 \;\de\\\nonumber
\spoiss{G_3}{\Phi_1'} & = & \frac{i}{\sqrt{2}} \left[ f'+Uf \right] X^- e_1^- \chid_1\;\de
                            +\; e_1^- h g' \chid_0 \;\de\\\nonumber
\spoiss{G_3}{\Phi_3'} & = & \frac{i}{\sqrt{2}} \left[ f'+Uf \right] X^- e_1^- \chi_1\;\de
                            -\; e_1^- h g' \chi_0 \;\de\,.
\eeqas

Several technical points in the calculation of the constraint algebra
may deserve some comments.  Only obtaining \eqref{G2G3} needs some
care, the other brackets are calculated rather straightforward using
the Poisson structure on phase space \eqref{poisson}. The tricky part
of \eqref{G2G3} is actually not the $C^{\al\be}$-term in the Dirac
bracket, but the bracket $\spoiss{G_2[\varphi]}{G_3[\psi]}$ and
therein the integrations by parts which have to be performed using
smeared constraints
\beqs
G_i[\varphi] = \int dx \; \varphi(x) G_i(x)\,.
\eeqs
The bracket itself reads with \eqref{G2}, \eqref{G3}
\beqa\nonumber
\spoiss{G_2[\varphi]}{G_3[\psi]} & = & \iint dx dz \varphi(x) \psi(z) \left( \spoiss{G_2^{h=0}(x)}{G_3^{h=0}(z)}\right.\\\label{sG2G3}
                                 &   & \hspace{-25mm}\left. + \spoiss{q^3(x)h(x)g(x)}{G_3^{h=0}(z)} - \spoiss{G_2^{h=0}(x)}{q²(z)h(z)g(z)}\right)\,.
\eeqa
Here $G^{h=0}$ denote the constraints with $h=0$,
\beqas
G_1^{h=0} & = & G_1^g = G_1 \\
G_2^{h=0} & = & G_2^g + \frac{i}{\sqrt{2}}f(X)(\chid_1 \lrpd{1} \chi_1)\\
G_3^{h=0} & = & G_3^g - \frac{i}{\sqrt{2}}f(X)(\chid_0 \lrpd{1} \chi_0)\,,
\eeqas
and
\beqs
\spoiss{G_2^{h=0}(x)}{G_3^{h=0}(z)} = - \sum\limits_{i=1}^3 \td{{\cal V}}{p_i} G_i^{g}\,.
\eeqs
The tricky parts are the second and third bracket in \eqref{sG2G3}%
,
\beqas
\spoiss{(q³h(p_1)g(\chib\chi))[\varphi]}{G_3^{h=0}[\psi]}
& = & \iint dx dz \varphi_x \psi_z g_x(\chib\chi) \poiss{q^3_xh_x(p_1)}{G_{3,z}^g} \\
& = & \iint dx dz \varphi_x \psi_z g_x(\chib\chi) \left[ (\partial_z \de(x-z))h_x(p_1) \right.\\
&   & \hspace{15mm}\left.- (q^1h-q^3p_3 h' -q^2p_2 U h)_x \;\de(x-z)  \right]\\
\spoiss{G_2^{h=0}[\varphi]}{(q²h(p_1)g(\chib\chi))[\psi]}
& = & \iint dx dz \varphi_x \psi_z g_z(\chib\chi) \poiss{G_{2,x}^g}{q^2_z h_z(p_1)} \\
& = & \iint dx dz \varphi_x \psi_z g_z(\chib\chi) \left[ (-\partial_x \de(x-z))h_z(p_1) \right.\\
&   & \hspace{15mm}\left.- (q^1h-q^2p_2 h' -q^3p_3 U h)_x\;\de(x-z)  \right]\,.
\eeqas
\beqas
& \Rightarrow & \hspace{12mm} \spoiss{(q³h(p_1)g(\chib\chi))[\varphi]}{G_3^{h=0}[\psi]} -
\spoiss{G_2^{h=0}[\varphi]}{(q²h(p_1)g(\chib\chi))[\psi]} \\
& & \hspace{3mm} = \hspace{5mm} \iint dx dz \varphi_x \psi_z [(\underbrace{(\partial_z \de(x-z))}\limits_{= -\partial_x \de(x-z)} g_x(\chib\chi) h_x(p_1) + (\partial_x \de(x-z)) g_z(\chib\chi) h_z(p_1)) \\
& & \hspace{35mm} \left. -g (q^2p_2-q³p_3)(h'(p_1)-U(p_1) h(p_1)) \de(x-z)\right]\\
& & \stackrel{int.p.p.}{=} \iint dx dz \varphi_x \psi_z \de(x-z)[ \partial_x(hg) - g(q²p_2-q³p_3)(h'-Uh) ]
\eeqas
One arrives at an expression for the graded Poisson bracket of $G_2$
and $G_3$ reading
\beqas\nonumber
\spoiss{G_2}{G_3 '} & = & \left[ -\td{{\cal V}}{p_i} G_i^g + \frac{i}{\sqrt{2}} f' [p_3(Q³\lrpd{x}Q^1)-p_2(Q²\lrpd{x}Q^0)]\right. \\
                    &   & \hspace{5mm}\left. + \partial_x(hg) - g(q²p_2-q³p_3)(h'-Uh)\right]\delta\,.
\eeqas

The $C^{\al\be}$-terms of the Dirac bracket are (with $\partial_xg =
g'\partial_x(\chib\chi)$, $\partial_x f = f' \partial_x p_1$ and
$p_2q²-p_3q³-\partial_x p_1 = -G_1$)
\beqa\nonumber
-\frac{i}{\sqrt{2}}[f'+Uf][p_3(Q³\lrpd{x}Q^1)-p_2(Q²\lrpd{x}Q^0)] - \frac{h}{f}f'g'\cdot(\chib\chi) G_1  - h(\partial_x g)\,.
\eeqa

With these results one obtains (omitting the $\de(x-x')$)
\beqa\nonumber
\dirac{G_2}{G_3 '} & = & -\td{\cal V}{p_1}G_1 
                         -\td{\cal V}{p_2}\left(G_2^g + \frac{i}{\sqrt{2}} f (Q³\lrpd{x}Q^1) + q³hg\right)\\\nonumber
                   &   & -\td{\cal V}{p_3}\left(G_3^g - \frac{i}{\sqrt{2}} f (Q³\lrpd{x}Q^1) - q²hg\right)\\\nonumber
                   &   & +\underbrace{\partial_x(hg)-h(\partial_x g) -gh'(\partial_x p_1)}\limits_{=0}
                         + g h' G_1\\\nonumber
                   &   & - \frac{h}{f}f' g'\cdot (\chib\chi)G_1\\\nonumber
                   & = & - \td{\cal V}{p_i}G_i + (gh' - \frac{h}{f}f' g'\cdot (\chib\chi))G_1\,.
\eeqa

\section{Dirac Brackets used in Sec. \ref{sec:witt}}\label{app:B2}

\beqas
	\dirac{G_1}{{q^1}'} & = & -\partial_1 \de \\
	\dirac{G_1}{{q^2}'} & = & q^2 \de \\
	\dirac{G_1}{{q^3}'} & = & -q^3 \de \\
	\dirac{G_2}{{q^1}'} & = & q^3\left[ \pd{\cal V}{p_1} - \left(h' g - \frac{f'}{f}h g' \cdot (\chib\chi)\right) \right] \de\\ 
	\dirac{G_2}{{q^2}'} & = & - \left[ \partial_1 + q^1 - q^3\pd{\cal V}{p_2}\right]\de \\
	\dirac{G_2}{{q^3}'} & = & q^3 \pd{\cal V}{p_3} \de\\
	\dirac{G_3}{{q^1}'} & = & -q^2\left[ \pd{\cal V}{p_1} - \left(h' g - \frac{f'}{f}h g' \cdot (\chib\chi)\right) \right] \de \\
	\dirac{G_3}{{q^2}'} & = & -q^2 \pd{\cal V}{p_2} \de \\
	\dirac{G_3}{{q^3}'} & = & - \left[ \partial_1 - q^1 + q^2\pd{\cal V}{p_3}\right]\de
\eeqas
\beqas
  \dirac{q^iG_i}{(q^iG_i)'} & = & -(\partial_1\de) q^i G_i \ \ \mathrm{(no\ summation\ over\ i)} \\
  \dirac{q^1G_1}{(q^2G_2)'} & = & -q^2q^3\left[ \pd{\cal V}{p_1} - \left(h' g - \frac{f'}{f}h g' \cdot (\chib\chi)\right) \right] G_1 \de\\
                         & = & - \dirac{q^1G_1}{(q^3G_3)'} = \dirac{q^2G_2}{(q^3G_3)'}
\eeqas

\section{Calculations from Sec. \ref{sec:BRST}}\label{app:B3}

The BRST charge $\Om$ is fermionic, thus the left hand side of
\eqref{eq:BRSTcancel} reads
$$\dirac{\Om^{(0)}}{\Om^{(1)'}} + \dirac{{\Om^{(0)}}'}{\Om^{(1)}}\,.$$
Thus if both Dirac brackets do not contain any terms proportional to
derivatives of the delta function this expression is just twice the
first bracket which reads
\begin{align*}
 &  \dirac{\Om^{(0)}}{\Om^{(1)'}}\\
= & \frac{1}{2}\dirac{c^iG_i}{(c^j c^k {C_{jk}}^l p_l^c)'}\\
= & \frac{1}{2} c^i\dirac{G_i}{{{C_{jk}}^l}'} {c^j}' {c^k}' {p_l^c}' - \frac{1}{2} c^j c^k {C_{jk}}^l G_l \de\,.
\end{align*}
The Dirac bracket $\dirac{G_i}{{{C_{jk}}^l}'}$ contains no $\de'$
contributions because of the following argument: There are three
Poisson brackets in the Dirac bracket \eqref{diracbracket}; the
bracket between the two arguments, and the brackets between the first
and second argument and the second class constraints, respectively.
The first one, $\spoiss{G_i}{{{C_{jk}}^l}'}$, does not give $\de'$
contributions because the constraints \eqref{G1}-\eqref{G3} contain
derivative terms $\partial_x p_i$ and $(Q^\dagger \lrpd{} Q)$, but the
structure functions ${C_{jk}}^l$ only depend on $p_i$ and $Q^\al$, cf.
eqs.  \eqref{GiGi} - \eqref{G2G3}. The same reasoning holds for
$\spoiss{\Phi_\al}{{{C_{jk}}^l}'}$, and the brackets
$\spoiss{G_i}{{\phi_\al}'}$ are given in App. \ref{app:B1}. Thus all
fields in the term containing $\dirac{G_i}{{{C_{jk}}^l}'}$ can be
taken at the same point,
$$c^i\dirac{G_i}{{{C_{jk}}^l}'} {c^j}' {c^k}' {p_l^c}' =
c^i\dirac{G_i}{{{C_{jk}}^l}} {c^j} {c^k} {p_l^c} \de(x-x')\,,$$
and the Dirac bracket between the constraints and the structure
functions can be calculated as if the fields where ordinary variables
in classical mechanics (i.e. without the delta functions in eqs.
\eqref{poisson} and \eqref{eq:gPBghost}). Terms quadratic in the
ghosts cancel, such that only the following Dirac brackets are
necessary for evaluating the first term in
$\dirac{\Om^{(0)}}{\Om^{(1)'}}$.
\beqa\label{eq:B1}
\dirac{G_2}{{C_{31}}^3} & = & \dirac{G_3}{{C_{12}}^2} = 0 \\\label{eq:B2}
\dirac{G_1}{{C_{23}}^1} & = & \spoiss{G_1}{{C_{23}}^1} = 0 \\\label{eq:B3}
\dirac{G_1}{{C_{23}}^2} & = & \spoiss{G_1}{{C_{23}}^2} = {C_{23}}^2 \de = - U(p_1) p_3 \de \\\label{eq:B4}
\dirac{G_1}{{C_{23}}^3} & = & \spoiss{G_1}{{C_{23}}^3} = - {C_{23}}^3 \de = U(p_1) p_2 \de\,.
\eeqa
Thus the left hand side of eq. \eqref{eq:BRSTcancel} reads
\eq{2 \dirac{\Om^{(0)}}{\Om^{(1)'}} = - c^j c^k {C_{jk}^l} G_l \de + 2 c^1 c^2 c^3 U(p_1) [ p_3^c p_2 - p_2^c p_3] \de\,.}{eq:BRSTcancelexpl}
It cancels eq. \eqref{eq:BRST00}, but produces a new contribution
which in turn is cancelled by the bracket
\beqas\nonumber
\dirac{\Om^{(1)}}{{\Om^{(1)}}'} & = & \frac{1}{4} \Big[ c^i c^j {C_{ij}}^k \spoiss{p_k^c}{(c^l c^m)'} {C_{lm}'}^n {p_n^c}'
                                                     + {C_{ij}}^k p_k^c \spoiss{c^i c^j}{{p_n^c}'} (c^l c^m)' {C_{lm}'}^n \\
                                & & \hspace{5mm} + c^i c^j p_k^c \underbrace{\dirac{{C_{ij}}^k}{{C_{lm}'}^n}}\limits_{= \dirac{{C_{ij}}^k}{{C_{lm}}^n} \de(x-x')} (c^l c^m p_n^c)' \Big]\\
                                & = & c^i c^j c^l {C_{ij}}^k {{C_{lk}}^n}p_n^c \de(x-x') + \frac{1}{4} \underbrace{c^i c^j c^l c^m}\limits_{= 0} p_k^c p_n^c \dirac{{C_{ij}}^k}{{C_{lm}}^n} \de(x-x') \\
                                & = & -2 c^1 c^2 c^3 U(p_1) [ p_3^c p_2 - p_2^c p_3] \de\,.
\eeqas
Here antisymmetry of the structure functions ${C_{ij}}^k = -
{C_{ji}}^k$ has been used in the penultimate step, and the explicit
form of the ${C_{ij}}^k$ (cf.
eqs.~\eqref{eq:structfunc}-\eqref{G2G3}) in the ultimate one. The
homological perturbation theory thus terminates at Yang-Mills level.

For calculating eqs.~\eqref{eq:B1}-\eqref{eq:B4} it is helpful to know
the following Poisson brackets of the second class constraints
eqs.~\eqref{Phi0}-\eqref{Phi3} with the structure functions,
\beqas
\spoiss{\Phi_\al}{{C_{12}'}^2} & = & \spoiss{\Phi_\al}{{C_{13}'}^3} = 0 \qquad \al = 0,1,2,3\\
\spoiss{\Phi_0}{{C_{23}'}^1} & = & \Big[-\frac{i}{\sqrt{2}} f(p_1)p_2 U' Q^2 + g' \left(h' - \frac{h}{f}f'\right)Q^3 - g'' \cdot (\chib\chi) \frac{h}{f}f' Q^3\Big]\de \\
\spoiss{\Phi_2}{{C_{23}'}^1} & = & \Big[-\frac{i}{\sqrt{2}} f(p_1)p_2 U' Q^0 - g' \left(h' - \frac{h}{f}f'\right)Q^1 + g'' \cdot (\chib\chi) \frac{h}{f}f' Q^1\Big]\de \\
\spoiss{\Phi_1}{{C_{23}'}^1} & = & \Big[+\frac{i}{\sqrt{2}} f(p_1)p_3 U' Q^3 + g' \left(h' - \frac{h}{f}f'\right)Q^2 - g'' \cdot (\chib\chi) \frac{h}{f}f' Q^2\Big]\de \\
\spoiss{\Phi_3}{{C_{23}'}^1} & = & \Big[+\frac{i}{\sqrt{2}} f(p_1)p_3 U' Q^1 - g' \left(h' - \frac{h}{f}f'\right)Q^0 + g'' \cdot (\chib\chi) \frac{h}{f}f' Q^0\Big]\de \\
\spoiss{\Phi_0}{{C_{23}'}^2} & = & - \frac{i}{\sqrt{2}} f U Q^2\de \qquad\qquad \spoiss{\Phi_1}{{C_{23}'}^3} = \frac{i}{\sqrt{2}} f U Q^3\de \\
\spoiss{\Phi_2}{{C_{23}'}^2} & = & - \frac{i}{\sqrt{2}} f U Q^0\de \qquad\qquad \spoiss{\Phi_3}{{C_{23}'}^3} = \frac{i}{\sqrt{2}} f U Q^1\de \\
\spoiss{\Phi_1}{{C_{23}'}^2} & = & \spoiss{\Phi_1}{{C_{23}'}^2} = \spoiss{\Phi_0}{{C_{23}'}^3} = \spoiss{\Phi_2}{{C_{23}'}^3} = 0\,.
\eeqas
%










\end{appendix}


\backmatter




\clearplaindoublepage
\lhead[\thepage]{\slshape \bibname}  
\rhead[\slshape \bibname]{\thepage}
\addcontentsline{toc}{chapter}{Bibliography}
\providecommand{\href}[2]{#2}\begingroup\raggedright\endgroup

\end{document}